\RequirePackage{amsthm} 

\documentclass[pdflatex,sn-mathphys-num]{sn-jnl}
\usepackage{graphicx}%
\usepackage{multirow}%
\usepackage{amsmath,amssymb,amsfonts}%
\usepackage{amsthm}%
\usepackage{mathrsfs}%
\usepackage[title]{appendix}%
\usepackage{xcolor}%
\usepackage{textcomp}%
\usepackage{manyfoot}%
\usepackage{booktabs}%
\usepackage{algorithm}%
\usepackage{algorithmicx}%
\usepackage{algpseudocode}%
\usepackage{listings}%
\usepackage{caption}
\usepackage{subcaption}
\usepackage{tabularx}
\usepackage{lmodern}
\usepackage{float}

\usepackage[ruled, lined, linesnumbered, algo2e, algosection]{algorithm2e}

\usepackage{xcolor}

\theoremstyle{thmstyleone}%
\theoremstyle{thmstyletwo}%

\theoremstyle{thmstylethree}%

\newcommand{\RR}{\mathbb{R}}

\newcommand{\eps}{\varepsilon}

\DeclareMathOperator*{\argmin}{arg\,min}

\begin{document}


\title{Node Splitting SVMs for Survival Trees Based on
an $L_2$-Regularized Dipole Splitting Criteria}


\author*[1]{\fnm{Aye Aye} \sur{Maung}}\email{ayeaye.maung@louisville.edu}

\author[2]{\fnm{Drew} \sur{Lazar}}\email{dmlazar@bsu.edu}

\author[1]{\fnm{Qi} \sur{Zheng}}\email{qi.zheng@louisville.edu}

\affil*[1]{\orgdiv{Department of Bioinformatics and Biostatistics}, \orgname{University of Louisville}, \orgaddress{\street{485 E. Gray St.}, \city{Louisville}, \postcode{40202}, \state{KY}, \country{USA}}}

\affil[2]{\orgdiv{Department of Mathematical Sciences}, \orgname{Ball State University}, \orgaddress{\street{Robert Bell Building, Room 465}, \city{Muncie}, \postcode{47306}, \state{IN}, \country{USA}}}


\abstract{This paper proposes a novel, node-splitting support vector machine (SVM) for creating survival trees. This approach is capable of non-linearly partitioning survival data which includes continuous, right-censored outcomes. Our method improves on an existing non-parametric method, which uses at most oblique splits to induce survival regression trees. In the prior work, these oblique splits were created via a non-SVM approach, by minimizing a piece-wise linear objective, called a dipole splitting criterion, constructed from pairs of covariates and their associated survival information. We extend this method by enabling splits from a general class of non-linear surfaces. We achieve this by ridge regularizing the dipole-splitting criterion to enable application of kernel methods in a manner analogous to classical SVMs. The ridge regularization provides robustness and can be tuned. Using various kernels, we induce both linear and non-linear survival trees to compare their sizes and predictive powers on real and simulated data sets. We compare traditional univariate log-rank splits, oblique splits using the original dipole-splitting criterion and a variety of non-linear splits enabled by our method.  In these tests, trees created by non-linear splits, using polynomial and Gaussian kernels show similar predictive power while often being of smaller sizes compared to trees created by univariate and oblique splits. This approach provides a novel and flexible array of survival trees that can be applied to diverse survival data sets. 
}


\maketitle
\section{Introduction} \label{sec:intro}

In the last 40 years, tree-based survival analysis models, which provide interpretable predictions without requiring the strong assumptions of parametric modeling techniques, have been developed. The first such model was introduced in~\cite{ciampi}, and further development was presented in~\cite{gordonolshen}, where nodes are split to attain the smallest amount of inter-node variability as measured by the Wasserstein metric. From there, many types of survival trees, mostly differing in the splitting criteria, were developed. These include univariate splits based on the log-rank statistic in~\cite{leblanccrowley}, splits based on likelihood ratio statistic and the assumption of constant hazards of the exponential model~\cite{bou2011review}, and completely non-parametric, oblique splits by hyperplanes in~\cite{kretowska}. In this paper, we build on the approach in~\cite{kretowska}.

The work in~\cite{kretowska} is an extension of a previous work by~\cite{bobrowskikretowski}, which introduced a non-parametric method for creating decision trees for non-survival data using oblique splits. The underlying idea in~\cite{bobrowskikretowski} is to classify pairs of covariate vectors, called dipoles, according to time differences. In~\cite{kretowska}, it was demonstrated how to account for right-censoring in this process. The aim is to find a hyperplane that splits many dipoles with large time differences, while splitting few dipoles with small time differences. This is achieved by associating one of four types of penalty functions to each dipole. A weighted sum of these penalty functions is known as a dipole splitting criteria function, and the optimal hyperplane is found through optimizing such a function.

An unresolved issue that arises in this process is determining which penalty functions to associate with each dipole. This issue is not adequately addressed in either~\cite{kretowska} or~\cite{bobrowskikretowski}. In particular, in~\cite{kretowska}, the notion of ``orientation'', which determines the assignment of penalty function to dipoles, is not rigorously specified. Resolving this issue is one of the auxiliary aims of this paper. We introduce and justify a rigorous definition of dipole orientation. Based on the definition, we detail an iterative reorientation algorithm to locate splits at each non-terminal node of the survival tree. Convergence of the algorithm is also demonstrated.

More broadly, the aim of the paper is to extend the oblique splits in~\cite{kretowska}, to allow splits by a large class of non-linear surfaces at each non-terminal node. This is achieved by introducing a ridge penalty term to the splitting problem of~\cite{kretowska} and then by applying the kernel method to the dual of the modified problem. The added ridge regularization term can also control overfitting, with the extent of this effect controlled by a tuning factor. This produces a non-linear, node-splitting SVM with a form most similar to support vector regression (SVR) machines of~\cite{Cortes1995}.

SVMs are versatile tools in statistical learning that have been applied well beyond their original use as classifiers of binary outcome data. For survival analysis, SVMs have been modified in a variety of ways to deal with both the continuous aspect of the outcome data and censoring. Survival SVM regression, due to \cite{SVMsurvivalregression}, seeks to estimate functions that best fit observed survival times using the covariates. On the other hand, survival SVM ranking, due to \cite{SVMsurvivalranking}, leverages the ordinal aspect of the observed survival times and aims to predict the optimal risk ranking among individuals instead. Both approaches benefit from the amenability of SVMs to non-linear fits of covariates via the kernel method. Hybrid SVMs such as \cite{SVMsurvivalhybrid}, that combine both regression and ranking approaches, also exist.

Our work differs from these approaches of \cite{SVMsurvivalregression, SVMsurvivalranking, SVMsurvivalhybrid}, in that our proposed SVM does not directly predict survival outcomes. Rather, it is a node-splitter for inducing survival regression trees. Its aim is to find partitions of non-terminal nodes such that pairs of survival times within each partition are as similar as possible, while pairs across partitions are as dissimilar as possible. Furthermore, in our approach, we iteratively reorientate and reoptimize the node-splitting SVM, using our definition of dipole orientation. This produces an improving sequence of splitting solutions, from each reorientated SVM, that converges to a final split of the non-terminal node. In this manner, our node-splitting method extends the algorithm of \cite{kretowska} in a novel fashion, by allowing for ridge-regularization and splitting by a wide class of nonlinear surfaces through the kernel method.


Using different kernels on our proposed node-splitting SVM, we compare oblique, quadratic and Gaussian splits to examine the advantages and disadvantages of non-linear splits over the linear splits of \cite{kretowska} in survival trees. These approaches are applied to and compared on real and simulated data sets. 

\section{Background}

\subsection{Survival Data} \label{sec:survdata}

We assume that $n$ observations, $(\mathbf{x}_j, t_j, \delta_j)_{j = 1}^n \subset \mathbb{R}^p \times \mathbb{R}_{+} \times \{0, 1\}$, are sampled from a random triple $(\mathbf{X}, T, \Delta)$.  $\mathbf{X}$ is a $p$-dimensional covariate vector. The observed survival time $T$ satisfies 
\[
T = \min(T_0, C),
\]
where $T_0$ is a positive random variable indicating survival time and $C$ is a positive random variable indicating right-censoring time. $T_0$ and $C$  are independent so that censoring is non-informative. $\Delta = I(T_0 < C)$ is a censoring indicator. 

\subsection{Survival Dipolar Criterion}\label{subsec:SDC} 

In~\cite{kretowska},~a non-parametric, tree-based method of partitioning the covariate space of right-censored data by hyperplanes was proposed, which extended to right-censored data the approach of \cite{bobrowskikretowski}. Both works classify pairs of covariate vectors,
\[
\{ (\mathbf{x}_i, \mathbf{x}_j) \}_{ 1 \leq i < j \leq n },
\]
known as \textit{dipoles}, according to their time-difference and censoring information. The approach in \cite{kretowska} is to first create a vector of pairwise time differences, $\Delta T$, from the pairs of survival information $\{(t_i, \delta_i), (t_j, \delta_j)\}_{1 
 \leq i < j \leq n}$ as below. 

\begin{enumerate}[(1)]
\item Call an index pair $(i, j)$ right-comparable if $\delta_k = 1$ where $k = \text{index}[\min(t_i, t_j)]$, i.e., if the smaller survival time is uncensored.
\item Letting $\mathcal{P}$ be the set of right-comparable pairs and $\ell = \lvert \mathcal{P} \rvert$, define $\Delta T \in \mathbb{R}^{\ell}$ to be the vector of pairwise differences $\lvert t_i - t_j \rvert$ of right-comparable pairs $(i, j)$.
\end{enumerate} 
Next, the aim roughly speaking, is to label dipoles with small time differences among $\Delta T$ as \textit{pure} and to label dipoles with large time differences among $\Delta T$ as \textit{mixed}. Percentile cutoffs  $0 < \zeta_1 < \zeta_2 < 1$ of $\Delta T$ are fixed to determine \emph{pure} and \emph{mixed} dipoles. The \textit{survival dipolar criterion} for $\{ (\mathbf{x}_i, \mathbf{x}_j) \}_{  1 \leq i < j \leq n }$ is then defined using the order statistics of $\Delta T$ as follows. 
\begin{enumerate}[(1)]
	\item $(\mathbf{x}_i, \mathbf{x}_j)$ is \emph{pure} if $\delta_i = \delta_j = 1$ and $\lvert t_i - t_j \rvert < \Delta T_{(\lfloor \zeta_1 \cdot L \rfloor)}$.
	\item $(\mathbf{x}_i, \mathbf{x}_j)$  is \emph{mixed} if $(i, j) \in \mathcal{P}$ and $\lvert t_i - t_j \rvert \geq \Delta T_{(\lfloor \zeta_2 \cdot L \rfloor)}$. 
	\item All other $(\mathbf{x}_i, \mathbf{x}_j)$ are \emph{neither}.
\end{enumerate}

\subsection{Dipole Penalty Functions} \label{sec:DPF}

With dipoles labeled by the survival dipolar criterion in Section~\ref{subsec:SDC}, \cite{kretowska} partitions the covariate space at non-terminal nodes in decision trees by hyperplanes. These hyperplanes are intended to split many mixed dipoles while splitting few pure dipoles. This is achieved by first associating special piecewise linear functions, called dipole penalty functions, to each dipole. Each dipole penalty function penalizes hyperplanes based on their position relative to the dipole. The dipole penalty functions are defined below in \eqref{eqs:mixedpenalty} and \eqref{eqs:purepenalty}, as simple combinations of hinge-loss functions \eqref{eqs:PLC}.

We identify hyperplanes
\[
H(\mathbf{v}) = \big\{(x_1,\ldots,x_p) \in \RR^p ; \mathbf{v} \cdot  \mathbf{z} = 0 \text{ where } \mathbf{z} = (1, x_1,\ldots,x_p) \big\}
\]
by their coefficient vectors $\mathbf{v} = (w_0, \mathbf{w}) \in \RR^{p + 1}$. To each covariate vector $\mathbf{x}_j$, we associate its augmented covariate vector $\mathbf{z}_j = (1, \mathbf{x}_j)$ and the hinge-loss functions $\varphi^+_j, \varphi^-_j : \mathbb{R}^{p+1} \to \mathbb{R}$, 
	\begin{equation}
		\varphi^+_j(\mathbf{v}) = \max\{0, \eps_j - \mathbf{v} \cdot   \mathbf{z}_j\} \text{ and } 
		\varphi^-_j(\mathbf{v}) = \max\{0, \eps_j + \mathbf{v} \cdot   \mathbf{z}_j\} \label{eqs:PLC}. 
	\end{equation} where $\eps_j > 0$ for $j = 1, \ldots, n$ are constants. Following \cite{bobrowskikretowski}, we assume uniform margins,
$\eps_j = \eps \text{ for } j = 1, \ldots, n \label{eqs:uniformmargins}$ and some fixed $\eps>0$.
\begin{figure}[H]
	\centering
        \includegraphics[scale=.25]{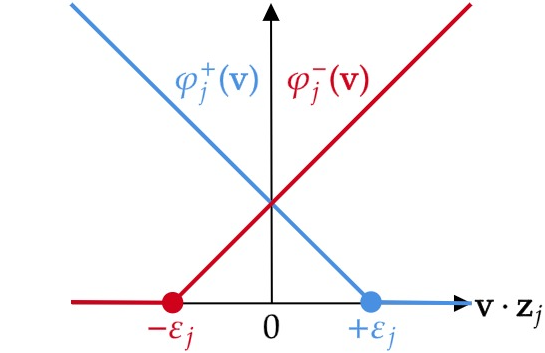}
        \caption{Hinge-loss Functions} \label{fig:1}
        \captionsetup{justification=centering}
\end{figure}

In \cite{kretowska}, the hinge-loss functions in \eqref{eqs:PLC} are combined to define two pairs of dipole penalty functions. The pair of functions
\begin{align}
	\varphi^{m^+}_{jk} = \varphi^+_j + \varphi^-_k \text{ and } \varphi^{m^-}_{jk} = \varphi^-_j + \varphi^+_k  \label{eqs:mixedpenalty}
\end{align} penalize mixed dipoles that remain unsplit. Given a dipole $(\mathbf{x}_j, \mathbf{x}_k)$, each of these functions is minimized by a $\mathbf{v}$ which splits the dipole. On the other hand, the pair of functions
\begin{align}
	\varphi^{p^+}_{jk} = \varphi^+_j + \varphi^+_k \text{ and } \varphi^{p^-}_{jk} = \varphi^-_j + \varphi^-_k \label{eqs:purepenalty}
\end{align} penalize the splitting of pure dipoles. Given a dipole $(\mathbf{x}_j, \mathbf{x}_k)$, each of these functions is minimized by a $\mathbf{v}$ which does not split the dipole.

\subsection{Dipole Splitting Functions} \label{sec:DCF}

 Using a weighted sum of penalty functions from~\eqref{eqs:mixedpenalty} and~\eqref{eqs:purepenalty}, \cite{kretowska} constructs an objective function that is minimized at a $\mathbf{v}$ for which many mixed dipoles are split and many pure dipoles are not split by $H(\mathbf{v})$. In the objective function, $\varphi^{m^+}_{jk}$ or $\varphi^{m^-}_{jk}$ from \eqref{eqs:mixedpenalty} are assigned to each mixed dipole, and $\varphi^{p^+}_{jk}$ or $\varphi^{p^-}_{jk}$ from \eqref{eqs:purepenalty} are assigned to each pure dipole.

While \cite{kretowska} states assignments are made according to the ``orientation" of the dipoles, each dipole and its two elements do not intrinsically possess ``orientation'' without respect to a particular $\mathbf{v}$.  The dipole splitting criterion function in  \cite{bobrowskikretowski} chooses $\varphi^{m^+}_{jk}$ for all mixed dipoles and chooses $\varphi^{p^+}_{jk}$ for all pure dipoles. In this paper, and as described in Section~\ref{sec:OCDPF},  functions from \eqref{eqs:mixedpenalty} and \eqref{eqs:purepenalty} are assigned to dipoles in a geometrically reasonable manner with respect to initial $\mathbf{v}_i$'s at each step of a recursive optimization algorithm. The dipole splitting function is then minimized with respect to those $\mathbf{v}_i$'s.

For now, we introduce the dipole splitting function presented in \cite{kretowska}. Let $I^{p^+}, I^{p^-}, I^{m^+}, I^{m^-}$ be the disjoint sets of pairs of indices of dipoles that are, respectively, pure with $\varphi^{p^+}_{jk}$ assigned, pure with $\varphi^{p^-}_{jk}$ assigned, mixed with $\varphi^{m^+}_{jk}$ assigned and mixed with $\varphi^{m^-}_{jk}$ assigned. Then the dipole splitting function $\Psi : \RR^{p + 1} \to \RR$ is

\begin{align}
	\Psi(\mathbf{v}) &= \sum_{\substack{(j, k) \\ \hspace{3.7pt} \in  I^{p^+}}} \alpha_{jk} \varphi^{p^+}_{jk}(\mathbf{v}) + \sum_{\substack{(j, k) \\ \hspace{3.7pt} \in I^{p^-}}} \alpha_{jk} \varphi^{p^-}_{jk}(\mathbf{v}) \nonumber \\ &+ \sum_{\substack{(j, k) \\ \hspace{3.7pt} \in I^{m^+}}} \alpha_{jk} \varphi^{m^+}_{jk}(\mathbf{v}) +\sum_{\substack{(j, k) \\ \hspace{3.7pt} \in I^{m^-}}} \alpha_{jk} \varphi^{m^-}_{jk}(\mathbf{v}). \label{eq:DPC_KRETOWSKA}
\end{align} The coefficients $\alpha_{jk} \in \mathbb{R}$ are nonnegative
   price factors of penalty functions. These can be fixed to prioritize splitting mixed dipoles or to prioritize not splitting pure dipoles depending on the relative sizes of the coefficients. Following \cite{bobrowskikretowski}, we set $\alpha_{jk} = 1$ for all $j,k$ to give equal priority.


\section{Methods} \label{sec:methods}

\subsection{Orientation and Choice of Dipole Penalty Functions} \label{sec:OCDPF}

In this section, we introduce a rigorous definition of dipole orientation with respect to a fixed hyperplane. We then specify how the penalty functions are assigned to the dipoles based on the orientation. Finally, we describe how the definition is used to construct orientated dipole splitting functions.

\subsubsection{Definition of Dipole Orientation}

As in Section~\ref{sec:DPF}, let $\mathbf{v}_i$ be a hyperplane coefficient vector and let $(\mathbf{z}_j, \mathbf{z}_k)$ be an augmented dipole.
\begin{enumerate}[(1)]
	\item Let $(\mathbf{z}_j, \mathbf{z}_k)$ be pure. It is \emph{positively orientated} if ${\mathbf{v}_i} \cdot   (\mathbf{z}_j + \mathbf{z}_k) \geq 0$, and in this case we assign $\varphi^{p^+}_{jk}$ to it. It is \emph{negatively oriented} if ${\mathbf{v}_i} \cdot   (\mathbf{z}_j + \mathbf{z}_k) < 0$, and in this case we assign $\varphi^{p^-}_{jk}$ to it.
	\item Let $(\mathbf{z}_j, \mathbf{z}_k)$ be mixed. It is \emph{positively oriented} if ${\mathbf{v}_i} \cdot   (\mathbf{z}_j - \mathbf{z}_k) \geq 0$, and in this case we assign $\varphi^{m^+}_{jk}$ to it. It is \emph{negatively oriented} if ${\mathbf{v}_i} \cdot   (\mathbf{z}_j - \mathbf{z}_k) < 0$, and in this case we assign $\varphi^{m^-}_{jk}$ to it.
\end{enumerate}

The index sets now depend on $\mathbf{v}_i$ and, accordingly, we write $I^{p^+}_{\mathbf{v}_i}, I^{p^-}_{\mathbf{v}_i}, I^{m^+}_{\mathbf{v}_i}, I^{m^-}_{\mathbf{v}_i}$. Through the index sets the dipole splitting functions also depend on the hyperplane and we get a $\mathbf{v}_i$-oriented dipole splitting function $\Psi_{\mathbf{v}_i} : \RR^{p + 1} \to \RR$:
\begin{align}
	\Psi_{\mathbf{v}_i}(\mathbf{v}) &= \sum_{\substack{(j, k) \\ \hspace{3.7pt} \in   I^{p^+}_{\mathbf{v}_i}}} \alpha_{jk} \varphi^{p^+}_{jk}(\mathbf{v}) + \sum_{\substack{(j, k) \\ \hspace{3.7pt} \in  I^{p^-}_{\mathbf{v}_i}}} \alpha_{jk} \varphi^{p^-}_{jk}(\mathbf{v}) \nonumber \\ &+ \sum_{\substack{(j, k) \\  \hspace{3.7pt} \in I^{m^+}_{\mathbf{v}_i}}} \alpha_{jk} \varphi^{m^+}_{jk}(\mathbf{v}) +\sum_{\substack{(j, k)   \\ \hspace{3.7pt} \in I^{m^-}_{\mathbf{v}_i}}} \alpha_{jk} \varphi^{m^-}_{jk}(\mathbf{v}). \label{eq:DPC}
\end{align}

\subsubsection{Orientation of Dipole Splitting Functions}

To justify our assignments of penalty functions based on orientation with respect to a particular $\mathbf{v}_i$, note the following:
\begin{align} 
	&\text{(a) If } {\mathbf{v}_i} \cdot  (\mathbf{z}_j + \mathbf{z}_k) \geq 0 \text{ then } \varphi^{p^-}_{jk}(\mathbf{v}_i) \geq \varphi^{p^+}_{jk}(\mathbf{v}_i) \nonumber \\
	&\text{(b) If }{\mathbf{v}_i} \cdot  (\mathbf{z}_j + \mathbf{z}_k) < 0 \text{ then } \varphi^{p^+}_{jk}(\mathbf{v}_i) > \varphi^{p^-}_{jk}(\mathbf{v}_i) \nonumber \\
	&\text{(c) If } {\mathbf{v}_i} \cdot  (\mathbf{z}_j - \mathbf{z}_k) \geq 0 \text{ then } \varphi^{m^-}_{jk}(\mathbf{v}_i) \geq \varphi^{m^+}_{jk}(\mathbf{v}_i) \nonumber \\
	&\text{(d) If } {\mathbf{v}_i} \cdot  (\mathbf{z}_j - \mathbf{z}_k) < 0 \text{ then } \varphi^{m^+}_{jk}(\mathbf{v}_i) > \varphi^{m^-}_{jk}(\mathbf{v}_i) \label{prop:penaltyjustify} 
\end{align}

Assuming uniform margins as in \eqref{eqs:uniformmargins}, we demonstrate (a) as follows.  
$$
{\mathbf{v}_i} \cdot  (\mathbf{z}_j + \mathbf{z}_k) \geq 0 \implies
\eps + {\mathbf{v}_i} \cdot  \mathbf{z}_j \geq \eps - {\mathbf{v}_i} \cdot   \mathbf{z}_k  \text{ and }
 \eps + {\mathbf{v}_i} \cdot  \mathbf{z}_k \geq \eps - {\mathbf{v}_i} \cdot   \mathbf{z}_j
$$
Thus,
\begin{align*}
&\varphi^-_j(\mathbf{v}_i) = \max\{0, \eps + {\mathbf{v}_i} \cdot  \mathbf{z}_j\} \geq \max\{0, \eps - {\mathbf{v}_i} \cdot  \mathbf{z}_k\} = \varphi^+_k(\mathbf{v}_i) \text{ and } \\
&\varphi^-_k(\mathbf{v}_i) = \max\{0, \eps + {\mathbf{v}_i} \cdot  \mathbf{z}_k\} \geq \max\{0, \eps - {\mathbf{v}_i} \cdot   \mathbf{z}_j\} = \varphi^+_j(\mathbf{v}_i)
\end{align*} and therefore,
$$
\varphi^{p^-}_{jk}(\mathbf{v}_i) = \varphi^-_j(\mathbf{v}_i) + \varphi^-_k(\mathbf{v}_i) \geq \varphi^+_j(\mathbf{v}_i) + \varphi^+_k(\mathbf{v}_i) = \varphi^{p^+}_{jk}(\mathbf{v}_i)
$$

\begin{figure}[H]
\centering
\begin{subfigure}{.5\textwidth}
  \centering
  \includegraphics[scale=.25]{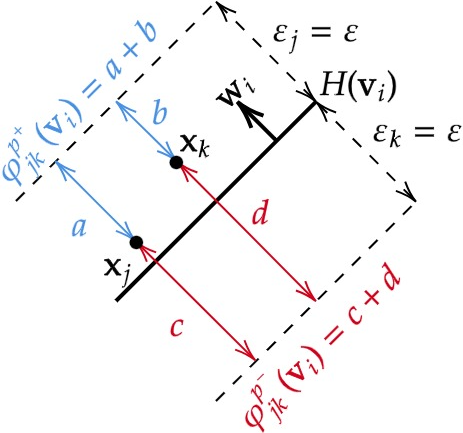}
  \caption{Pure, Positively Orientated}
  \label{fig:purepositive}
\end{subfigure}%
\begin{subfigure}{.5\textwidth}
  \centering
  \includegraphics[scale=.25]{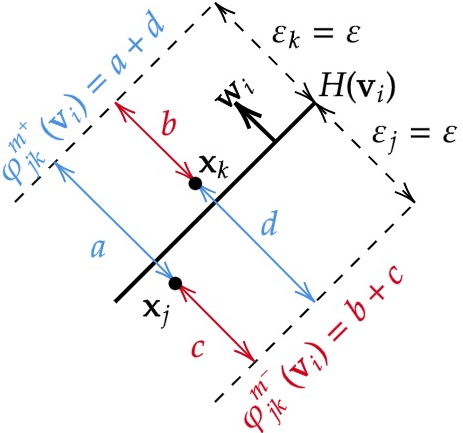}
  \caption{Mixed, Negatively Orientated}
  \label{fig:mixednegative}
\end{subfigure}
\caption{Penalties for Pure and Mixed Dipoles}
\label{fig:dipolepenalties}
\end{figure}

Hence, assigning $\varphi^{p^+}_{jk}$ rather than $\varphi^{p^-}_{jk}$ to a positively-oriented pure dipole doesn't penalize an initial splitting hyperplane $\mathbf{v}_i$ unnecessarily. We see examples geometrically in Figure~\ref{fig:dipolepenalties}, with $\lVert \mathbf{w}_i \rVert = 1$ assumed for simplicity. Figure~\ref{fig:purepositive} shows a typical positively-oriented pure dipole $(\mathbf{x}_j,\mathbf{x}_k)$.  This dipole is assessed a penalty of the sum of the distances to the closer margins by $\varphi^{p^+}_{jk}$ rather than a penalty of the sum of the distances to the further margins by $\varphi^{p^-}_{jk}$. A similar situation is shown in Figure~\ref{fig:mixednegative}, with a typical negatively-oriented mixed dipole $(\mathbf{x}_j,\mathbf{x}_k)$.

Thus, if we have an initial guess $\mathbf{v}_0$ of a split, and if we have a method that can use $\Psi_{\mathbf{v}_0}$ to get a better split $\mathbf{v}_1$, we can reorientate with respect to the better split and guarantee
\begin{gather}
\Psi_{\mathbf{v}_0}(\mathbf{v}_1) \geq \Psi_{\mathbf{v}_1}(\mathbf{v}_1) \label{dcfreorient}
\end{gather} due to~\eqref{prop:penaltyjustify}. This enables us to continue the process by using $\Psi_{\mathbf{v}_1}$ to get an even better split $\mathbf{v}_2$, and so on. The method to use an existing $\Psi_{\mathbf{v}_i}$ to get the better split $\mathbf{v}_{i + 1}$ will be described in the next Section~\ref{sec:optdcf}. Altogether, this will culminate in Section~\ref{sec:ROA} with an iterative algorithm to get progressively better splits by starting from an initial guess.

\subsection{Optimization of \texorpdfstring{$L_2$}{Ridge}-Regularized Dipole Splitting Functions} \label{sec:optdcf}

We take a different approach from both \cite{kretowska} and \cite{bobrowskikretowski} to finding splitting surfaces. In \cite{kretowska} and \cite{bobrowskikretowski}, splitting hyperplanes were found by directly minimizing the dipole splitting criterion \eqref{eq:DPC_KRETOWSKA}. In our case, we first opt to add a ridge penalty term to the dipole splitting criterion. Such penalty terms are seen in classification problems that use soft margin SVMs \cite{Cortes1995}. As in soft margin SVMs, this penalty term serves dual purposes. Firstly, it can be used to limit the complexity of splits to avoid overfitting. Secondly, this penalty term enables kernelization of the dual problem. Thus, it permits access to a large class of non-linear splits that were not available in \cite{kretowska}. 

\subsubsection{The Primal Problem}

Using~\eqref{eqs:purepenalty}, we express~\eqref{eq:DPC} solely in terms of the functions $\varphi^+_j, \varphi^-_j$. Indeed, if we define the coefficients $\beta^+_{j, \mathbf{v}_i}, \beta^-_{j, \mathbf{v}_i} \geq 0$ for $j = 1, \ldots, n$ as
\begin{align*}
\beta^+_{j, \mathbf{v}_i} &= \sum_{ \substack{k : (j, k) \\ \in I^{p^+}_{\mathbf{v}_i}} } \alpha_{jk} + \sum_{ \substack{k : (k, j) \\ \in I^{p^+}_{\mathbf{v}_i}} } \alpha_{kj} + \sum_{ \substack{k : (j, k) \\ \in I^{m^+}_{\mathbf{v}_i}} } \alpha_{jk} + \sum_{ \substack{k : (k, j) \\ \in I^{m^-}_{\mathbf{v}_i}} } \alpha_{kj} \\
\beta^-_{j, \mathbf{v}_i} &= \sum_{ \substack{k : (j, k) \\ \in I^{p^-}_{\mathbf{v}_i}} } \alpha_{jk} + \sum_{ \substack{k : (k, j) \\ \in I^{p^-}_{\mathbf{v}_i}} } \alpha_{kj}
	+ \sum_{ \substack{k : (j, k) \\ \in I^{m^-}_{\mathbf{v}_i}} } \alpha_{jk} + \sum_{ \substack{k : (k, j) \\ \in I^{m^+}_{\mathbf{v}_i}} } \alpha_{kj}
\end{align*} then \eqref{eq:DPC} can be rewritten as
\begin{align*}
\Psi_{\mathbf{v}_i}(\mathbf{v}) &= \sum_{j} \Big(\beta^+_{j, \mathbf{v}_i}\max\big\{0, \eps_j - \mathbf{v} \cdot  \mathbf{z}_j\big\} + \beta^-_{j, \mathbf{v}_i}\max\big\{0, \eps_j + \mathbf{v} \cdot  \mathbf{z}_j\big\}\Big)
\end{align*} for $\mathbf{v} \in \RR^{p + 1}$. Denoting $\mathbf{v} = (w_0, \mathbf{w})$, we add a ridge term to $\Psi_{\mathbf{v}_i}(\mathbf{v})$ to penalize large slopes $\mathbf{w}$ relative to the intercept $w_0$. This gives us a ridge-regularized dipole splitting function 
\begin{equation} 
\begin{gathered} 
    \Psi_{\mathbf{v}_i} (\mathbf{v};\kappa) = \frac{1}{2} \mathbf{w} \cdot  \mathbf{w}
    + \kappa \sum_{j} \Big(\beta^+_{j, \mathbf{v}_i}\max\big\{0, \eps_j - \mathbf{v} \cdot  \mathbf{z}_j\big\}  \\ + \beta^-_{j, \mathbf{v}_i}\max\big\{0, \eps_j + \mathbf{v} \cdot  \mathbf{z}_j\big\}\Big). \label{ridgedcf}
\end{gathered}
\end{equation}
The tuning parameter $\kappa > 0$ controls the extent of the effect of ridge-regularization. The minimization problem
$$
\argmin_{\mathbf{v} \in \RR^{p + 1}} \Psi_{\mathbf{v}_i}(\mathbf{v};\kappa)
$$ is equivalent to the convex quadratic program
\begin{equation}
\begin{gathered}
	\min_{w_0, \mathbf{w}, \boldsymbol{\xi}} \frac{1}{2}\mathbf{w} \cdot \mathbf{w} + \kappa\sum_i (\xi_j^+ + \xi_j^-)  \\
	\beta^+_{j, \mathbf{v}_i}(\eps_j - \mathbf{w} \cdot \mathbf{x}_j - w_0) \leq \xi_j^+ \\
	\beta^-_{j, \mathbf{v}_i}(\eps_j + \mathbf{w} \cdot \mathbf{x}_j + w_0) \leq \xi_j^- \\
	\xi_j^+, \xi_j^- \geq 0 \label{primal}
\end{gathered}
\end{equation}
for $j = 1, \ldots, n$. Minimization is done over the hyperplane variable $\mathbf{v} \in \RR^{p + 1}$ and the slack variables $\boldsymbol{\xi}^+, \boldsymbol{\xi}^- \in \RR^n$. 

\subsubsection{The Dual Problem} \label{sec:dualoptproblem}

Because its objective function is convex quadratic and the constraints are affine, the primal problem in~\eqref{primal} satisfies strong duality via Slater's condition. As a result, solutions of the primal problem and its dual counterpart necessarily coincide as in~\cite{Boyd_Vandenberghe_2004}. In turn, the formulation of the dual problem permits kernelization. The Lagrangian is
\begin{gather*}
\mathcal{L}(w_0, \mathbf{w}, \boldsymbol{\xi}, \boldsymbol{\gamma}, \boldsymbol{\tau}) = \frac{1}{2}\mathbf{w} \cdot \mathbf{w} + \kappa\sum_j (\xi_j^+ + \xi_j^-) + \sum_j \gamma_j^+[\beta^+_{j, \mathbf{v}_i}(\varepsilon_j - \mathbf{w} \cdot \mathbf{x}_j - w_0) - \xi_j^+] 
 \\ + \sum_j  \gamma_j^-[\beta^-_{j, \mathbf{v}_i}(\varepsilon_j + \mathbf{w} \cdot \mathbf{x}_j + w_0) - \xi_j^-]
- \sum_j \tau_j^+ \xi_j^+ - \sum_j \tau_j^- \xi_j^-
\end{gather*} where $\boldsymbol{\gamma}^+, \boldsymbol{\gamma}^-, \boldsymbol{\tau}^+, \boldsymbol{\tau}^- \in \mathbb{R}^n$ are non-negative Lagrange multipliers.

Setting gradients of the Lagrangian to zero, substituting solutions into the Lagragian, considering the Karush-Kuhn-Tucker (KKT) conditions \cite{Boyd_Vandenberghe_2004} and suitable reparametrizations, produces the dual problem below. 
\begin{equation}
\begin{gathered}
	\max_{\boldsymbol{\mu}} \sum_j (\mu_j^+ + \mu_j^-)\varepsilon_j - \frac{1}{2}\sum_{j, k} (\mu_j^+ - \mu_j^-)(\mu_k^+ - \mu_k^-)\mathbf{x}_j \cdot \mathbf{x}_k \\
	 \sum_j (\mu_j^+ -  \mu_j^-) = 0 \\
	\kappa\beta^+_{j, \mathbf{v}_i} \geq \mu_j^+ \geq 0, \qquad \kappa\beta^-_{j, \mathbf{v}_i} \geq \mu_j^- \geq 0 \label{dual}
\end{gathered}
\end{equation} for $j = 1, \ldots, n$. The dual problem is optimized over the variables $\boldsymbol{\mu}^{+}, \boldsymbol{\mu}^{-} \in \mathbb{R}^n$.

\subsection{Kernelization of the Dual Problem for Non-linear Splits} \label{sec:dualkernel}

We implement higher-order splits of data at non-terminal nodes of our decision trees. Explicit representation of higher-order surfaces through covariate augmentation is computationally expensive and often infeasible. However, the objective function of our dual problem in~\eqref{dual} only involves the inner product of covariates and, as in section \ref{sec:dualoptproblem}, our dual problem offers identical solutions to our primal problem. Thus, we can implement the kernel trick of SVMs \cite{Cortes1995} to split the feature space non-linearly. 

 We replace the inner product in~\eqref{dual} with a symmetric, positive semi-definite function $K : \RR^p \times \RR^p \to \RR$ called the \textit{kernel function}. 
 Mercer's Theorem~\cite{Cortes1995} guarantees the existence a feature map or embedding: $\phi : \RR^p \to \mathcal{V}$, where $(\mathcal{V}, \langle \cdot, \cdot \rangle)$ is a real, higher dimensional inner product space, satisfying $K(\mathbf{u}, \mathbf{v}) = \langle \phi(\mathbf{u}), \phi(\mathbf{v}) \rangle$ for all $\mathbf{u}, \mathbf{v}$. The kernel can be used to replace computationally expensive or infeasible computations involving $\phi$ and elements of $\mathcal{V}$. 

\subsubsection{The Kernelized Dual Problem} \label{sec:kernelized}

After expansion into the feature space $\mathcal{V}$, we no longer optimize the ridge-regularized dipole splitting function \eqref{ridgedcf}. Instead, we optimize its feature expanded counterpart $\widetilde{\Psi}_{\boldsymbol{\upsilon}_i}(\cdot, \cdot \ ; \kappa) : \RR \times \mathcal{V} \to \RR$
\begin{gather}
    \widetilde{\Psi}_{\boldsymbol{\upsilon}_i}(\omega_0, \boldsymbol{\omega}\ ; \kappa) = \frac{1}{2} \langle \boldsymbol{\omega} ,  \boldsymbol{\omega} \rangle
    + \kappa \sum_{j} \Big(\beta^+_{j, \boldsymbol{\upsilon}_i}\max\big\{0, \eps_j - \omega_0 - \langle \boldsymbol{\omega},  \phi(\mathbf{x}_j) \rangle \big\} \nonumber \\
    + \beta^-_{j, \boldsymbol{\upsilon}_i}\max\big\{0, \eps_j + \omega_0 + \langle \boldsymbol{\omega},  \phi(\mathbf{x}_j) \rangle\big\}\Big). \label{ridgedcfex}
\end{gather} Dipole orientation is now done with respect to a surface vector $\boldsymbol{\upsilon}_i \in \RR \times \mathcal{V}$ instead of a hyperplane vector. This is possible as the definitions and results of Section~\ref{sec:OCDPF} are preserved by replacing the inner product terms $\mathbf{v}_i \cdot \mathbf{z}_j$ with feature-expanded counterparts $\omega_{0, i} + \langle \boldsymbol{\omega}_i,  \phi(\mathbf{x}_j) \rangle$.

After replacement of the inner product in~\eqref{dual} with the kernel, the dual problem  becomes
\begin{gather}
	\max_{\boldsymbol{\mu}} \sum_j (\mu_j^+ + \mu_j^-)\eps_j - \frac{1}{2}\sum_{j, k} (\mu_j^+ - \mu_j^-)(\mu_k^+ - \mu_k^-) K(\mathbf{x}_j, \mathbf{x}_k) \nonumber \\
	 \sum_j (\mu_j^+ -  \mu_j^-) = 0 \nonumber \\
	\kappa\beta^+_{j, \boldsymbol{\upsilon}_i} \geq \mu_j^+ \geq 0 \qquad \kappa\beta^-_{j, \boldsymbol{\upsilon}_i} \geq \mu_j^- \geq 0 \label{kernelizeddual}
\end{gather} for $j = 1, \ldots, n$. The problem is optimized over the variables $\boldsymbol{\mu}^{+}, \boldsymbol{\mu}^{-} \in \RR^n$. This is the dual of the feature-expanded primal problem 
$$
\argmin_{\boldsymbol{\upsilon} \in \RR \times \mathcal{V}} \widetilde{\Psi}_{\boldsymbol{\upsilon}_i}(\omega_0, \boldsymbol{\omega}\ ; \kappa).
$$ Quadratic programming algorithms capable of handling positive semi-definite programming can be used to optimize this problem. In this paper, the operator-splitting algorithm in~\cite{osqpPaper} is used for this purpose. Along with the dual solutions, $\boldsymbol{\mu}_*^+, \boldsymbol{\mu}_*^-$, the primal intercept, $\omega_{0,*}$ is also estimated. 

\subsubsection{Orienting Dipoles using Dual Computations} \label{sec:dualcomp}
Given dual solutions, $\boldsymbol{\mu}_\ast^+, \boldsymbol{\mu}_\ast^-$ we can do computations with the primal solution $\boldsymbol{\omega}_\ast \in \mathcal{V}$ and the feature map $\phi$, without explicitly expressing either of them. Leveraging the KKT conditions of section~\ref{sec:dualoptproblem}, these computations rely on the relation 
\begin{gather}
    \langle \boldsymbol{\omega}_\ast, \phi(\mathbf{x}) \rangle = \sum_j (\mu_{j, \ast}^+ - \mu_{j, \ast}^-) K(\mathbf{x}_j, \mathbf{x}) \label{kernrel1}
\end{gather}
for any $\mathbf{x} \in \mathbb{R}^p$.
In particular, using \eqref{kernrel1}, we can replace inner product terms involving $\boldsymbol{\omega}_\ast$ and $\phi$ and we can orientate dipoles with respect to $\boldsymbol{\upsilon}_\ast$ using only its dual counterparts $\boldsymbol{\mu}_\ast^+, \boldsymbol{\mu}_\ast^-$ and its intercept $\omega_{0, \ast}$. Indeed, as remarked in Section \ref{sec:kernelized}, we can reformulate Section \ref{sec:OCDPF} by replacing terms such as $\mathbf{v}_i \cdot \mathbf{z}_j$ with feature-expanded counterparts $\omega_{0, i} + \langle \boldsymbol{\omega}_i,  \phi(\mathbf{x}_j) \rangle$. The inner product in the latter term can then be replaced with relation \eqref{kernrel1} to get a term involving only $\boldsymbol{\mu}_\ast^+, \boldsymbol{\mu}_\ast^-$ and $\omega_{0, \ast}$.

\subsection{Iterative Algorithm for Splitting Surfaces} \label{sec:ROA}

We use the orientation of dipoles of Section~\ref{sec:OCDPF}, and our optimization routine of Section~\ref{sec:kernelized}, to specify our algorithm to find splitting surfaces.  The algorithm is initialized with a univariate hyperplane, $\boldsymbol{\upsilon}_0$, in the original feature space, i.e., $\boldsymbol{\upsilon}_0 = (\omega_{0, 0}, \boldsymbol{\omega}_0) \in \mathbb{R} \times \mathbb{R}^{p}$ where $\boldsymbol{\omega}_0$ is a standard basis vector in $\mathbb{R}^p$ corresponding to a covariate. We choose $\omega_{0, 0}$ to be the negative of the median of this covariate. Since $\boldsymbol{\upsilon}_0$ is a hyperplane, the initial orientation of dipoles can be done as in Section~\ref{sec:OCDPF}. Our initializing  $\boldsymbol{\upsilon}_0$ is chosen to minimize~\eqref{ridgedcf} among the set of univariate hyperplanes through the medians of covariates after orientation.  

Next, we optimize \eqref{kernelizeddual} to get dual solutions
$(\boldsymbol{\mu}_1^+, \boldsymbol{\mu}_1^-, \omega_{0, 1})$ corresponding to 
$$
\boldsymbol{\upsilon}_1 = \argmin\limits_{\boldsymbol{\upsilon} \in \mathbb{R} 
\times \mathcal{V}} \widetilde{\Psi}_{\boldsymbol{\upsilon}_0}(\boldsymbol{\upsilon}  \ ; \kappa).
$$ Then $\widetilde{\Psi}_{\boldsymbol{\upsilon}_0}(\boldsymbol{\upsilon}_0  \ ; \kappa) \geq \widetilde{\Psi}_{\boldsymbol{\upsilon}_0}(\boldsymbol{\upsilon}_1  \ ; \kappa)$, so that $\boldsymbol{\upsilon}_1$ is a better candidate for the splitting surface than the initial $\boldsymbol{\upsilon}_0$. Hence, we can form a better feature-expanded dipole splitting criterion function, $\widetilde{\Psi}_{\boldsymbol{\upsilon}_1}$, by orienting the dipoles with respect to $\boldsymbol{\upsilon}_1$. Note that this is done using the dual solutions $(\boldsymbol{\mu}_1^+, \boldsymbol{\mu}_1^-, \omega_{0, 1})$ only, as remarked in Section~\ref{sec:dualcomp}. Thus, similar to what we saw in~\eqref{dcfreorient},
$
\widetilde{\Psi}_{\boldsymbol{\upsilon}_0}(\boldsymbol{\upsilon}_1) \geq \widetilde{\Psi}_{\boldsymbol{\upsilon}_1}(\boldsymbol{\upsilon}_1)
$ so that
\[
\widetilde{\Psi}_{\boldsymbol{\upsilon}_0}(\boldsymbol{\upsilon}_1 \ ; \kappa) =  \frac{1}{2} \langle \boldsymbol{\omega}_1, \boldsymbol{\omega}_1 \rangle + \widetilde{\Psi}_{\boldsymbol{\upsilon}_0}(\boldsymbol{\upsilon}_1) \geq \frac{1}{2} \langle \boldsymbol{\omega}_1, \boldsymbol{\omega}_1 \rangle + \widetilde{\Psi}_{\boldsymbol{\upsilon}_1}(\boldsymbol{\upsilon}_1) = \widetilde{\Psi}_{\boldsymbol{\upsilon}_1}(\boldsymbol{\upsilon}_1 \ ; \kappa). 
\] Continuing in this fashion, we get a sequence of dual solutions
$$
(\boldsymbol{\mu}_1^+, \boldsymbol{\mu}_1^-, \omega_{0, 1}), (\boldsymbol{\mu}_2^+, \boldsymbol{\mu}_2^-, \omega_{0, 2}), (\boldsymbol{\mu}_3^+, \boldsymbol{\mu}_3^-, \omega_{0, 3}), \ldots
$$ corresponding to the improving splitting surface vector candidates
$$
\boldsymbol{\upsilon}_1 = \argmin_{\boldsymbol{\upsilon} \in \mathbb{R} \times \mathcal{V}} \widetilde{\Psi}_{\boldsymbol{\upsilon}_0}(\boldsymbol{\upsilon}\ ; \kappa), \boldsymbol{\upsilon}_2 = \argmin_{\boldsymbol{\upsilon} \in \mathbb{R} \times \mathcal{V}} \widetilde{\Psi}_{\boldsymbol{\upsilon}_1}(\boldsymbol{\upsilon}\ ; \kappa), \boldsymbol{\upsilon}_3 = \argmin_{\boldsymbol{\upsilon} \in \mathbb{R} \times \mathcal{V}} \widetilde{\Psi}_{\boldsymbol{\upsilon}_2}(\boldsymbol{\upsilon}\ ; \kappa), \ldots
$$ and these satisfy
\begin{gather}
	\widetilde{\Psi}_{\boldsymbol{\upsilon}_0}(\boldsymbol{\upsilon}_0\ ; \kappa) \geq \widetilde{\Psi}_{\boldsymbol{\upsilon}_1}(\boldsymbol{\upsilon}_1\ ; \kappa) \geq \widetilde{\Psi}_{\boldsymbol{\upsilon}_2}(\boldsymbol{\upsilon}_2\ ; \kappa) \geq  \widetilde{\Psi}_{\boldsymbol{\upsilon}_3}(\boldsymbol{\upsilon}_3\ ; \kappa) \geq \cdots \label{eq:DPC_evalseq}.
\end{gather} The decreasing sequence~\eqref{eq:DPC_evalseq} converges as it is nonnegative.

Hence, we get the convergent algorithm \eqref{algo:roo} to arrive at a dual solution $(\boldsymbol{\mu}_\ast^+, \boldsymbol{\mu}_\ast^-, \omega_{0, \ast})$ corresponding to an improved splitting surface $\boldsymbol{\upsilon}_\ast$ of our data.

\SetKwRepeat{Do}{do}{while}

\begin{algorithm2e}
\DontPrintSemicolon
\KwData{initial hyperplane guess $\boldsymbol{\upsilon}_0 \in \RR \times \mathcal{V}$ and tolerance $\tau > 0$}
\KwResult{dual solution $(\boldsymbol{\mu}_\ast^+, \boldsymbol{\mu}_\ast^-, \omega_{0, \ast}) \in \RR^n \times \RR^n \times \RR$ corresponding to a splitting surface vector $\boldsymbol{\upsilon}_\ast \in \RR \times \mathcal{V}$}
\Begin{
    $\widetilde{\Psi}_{\boldsymbol{\upsilon}_0}(\cdot \ ; \kappa) \Leftarrow \text{ orient dipoles with } \boldsymbol{\upsilon}_0$\;
    $(\boldsymbol{\mu}_\ast^+, \boldsymbol{\mu}_\ast^-, \omega_{0, \ast}) \Leftarrow \text{dual of } \boldsymbol{\upsilon}_\ast = \argmin\limits_{\boldsymbol{\upsilon} \in \RR \times \mathcal{V}} \widetilde{\Psi}_{\boldsymbol{\upsilon}_0}(\boldsymbol{\upsilon}\ ; \kappa)$\;
    \Do{$\lvert \widetilde{\Psi}_{\boldsymbol{\upsilon}_\ast}(\boldsymbol{\upsilon}_\ast\ ; \kappa) -  \widetilde{\Psi}_{\boldsymbol{\upsilon}'}(\boldsymbol{\upsilon}'\ ; \kappa)\rvert > \tau$}{
    		$(\boldsymbol{\mu}_{'}^+, \boldsymbol{\mu}_{'}^-, \omega_{0}') \Leftarrow (\boldsymbol{\mu}_\ast^+, \boldsymbol{\mu}_\ast^-, \omega_{0, \ast})$\;
            $\widetilde{\Psi}_{\boldsymbol{\upsilon}'}(\cdot \ ; \kappa) \Leftarrow \text{ orient dipoles with }(\boldsymbol{\mu}_{'}^+, \boldsymbol{\mu}_{'}^-, \omega_{0}')$\;
            $(\boldsymbol{\mu}_\ast^+, \boldsymbol{\mu}_\ast^-, \omega_{0, \ast}) \Leftarrow \text{dual of } \boldsymbol{\upsilon}_\ast = \argmin\limits_{\boldsymbol{\upsilon} \in \RR \times \mathcal{V}} \widetilde{\Psi}_{\boldsymbol{\upsilon}'}(\boldsymbol{\upsilon}\ ; \kappa)$\;
            $\widetilde{\Psi}_{\boldsymbol{\upsilon}_\ast}(\cdot \ ; \kappa) \Leftarrow \text{ orient dipoles with }(\boldsymbol{\mu}_\ast^+, \boldsymbol{\mu}_\ast^-, \omega_{0, \ast})$\;
    }
}
\caption{Recursive Reorientation and Optimization} \label{algo:roo}
\end{algorithm2e}

\subsection{Growth of Trees and Pruning} \label{sec:prune} Our trees are grown by creating branches through recursive splitting of training sets. A terminal node is reached if either the node contains less than a fixed number of observations, or if the node contains all pure dipoles.  
As in~\cite{leblanc1993survival},  pruning is then done of trees grown from training sets by successively removing branches that have the smallest ratio
\begin{equation} \label{prunratio}
g(h) = \frac{G(\mathcal{T}_h)}{\lvert S_h \rvert}
\end{equation}
where $G(\mathcal{T}_h)$ is the sum of log-rank statistics at and below the root of the branch $h$, and $\lvert S_h \rvert$ is the number of non-terminal nodes at or below the root of the branch $h$. Letting $\mathcal{T}_0$ be our initial tree and $\mathcal{T}_m$ its root node, with $\mathcal{T}_{k+1}$ the subtree of $\mathcal{T}_k$ obtained by pruning using \eqref{prunratio}, the series of subtrees
\begin{equation} \mathcal{T}_m \prec \ldots \prec \mathcal{T}_1 \prec \mathcal{T}_0
\label{recsubtrees}
\end{equation} 
is obtained.
Let 
\begin{equation*}
    \alpha_i = \min_{h \in \mathcal{T}_{i-1}} \frac{G(\mathcal{T}_h)}{\lvert S_h \rvert} \text{ for } i=1,\ldots,m.  
\end{equation*}
In~\cite{leblanc1993survival}, it is shown that $\alpha_{m}>\ldots>\alpha_1$ and that for any $\alpha_k \le \alpha < \alpha_{k+1}$, $\mathcal{T}_k$ maximizes
\begin{equation}
G_\alpha(\mathcal{T}') = G(\mathcal{T}') - \alpha \lvert S' \rvert
\label{eq:splitcomplexity}
\end{equation}
over subtrees $\mathcal{T}'$ of $\mathcal{T}$. $G_\alpha(\mathcal{T}')$ is the split complexity of the tree and accounts for the trade-off between the ``progonostic structure'' of the tree reflected in the sum of its logrank statistics and the complexity of the tree in the number of its non-terminal nodes. 

Given training and validation samples, we can choose a subtree by first setting $2 \leq \alpha_c \leq 4$ (upper percentiles from the $\chi^2_1$ distribution). Then we choose from~\eqref{recsubtrees} the subtree that maximizes $G_{\alpha_c}$ when validation data is sent down each of these subtrees. For bias correction,
bootstrapping can be carried out to select from~\eqref{recsubtrees}, and  we implement this  procedure.

\section{Results} \label{sec:results}
 Here we discuss the results of validating our method from Section~\ref{sec:methods}. In Section~\ref{sec:COK}, we detail the specific kernels from Section~\ref{sec:dualkernel}, used for our models. In Section~\ref{sec:evalofmodels}, we detail assessment measures.  We apply the models to simulated data in Section~\ref{sec:simresults}, and to real data in Section~\ref{sec:appresults}. In Section~\ref{sec:detsplits}, we illustrate the ability of the algorithm of Section~\ref{sec:ROA} to detect splits, by various hazards, of simulated survival data.  In Section~\ref{sec:modevalsimdata}, we evaluate survival trees, created by splits through our algorithm of Section~\ref{sec:ROA}, on data generated by simulation.  In Section~\ref{sec:remisisondata}, we illustrate the characteristics of the models through their application to a small data set. In Section~\ref{sec:addrealdata}, we then apply our models to several larger real data sets and consider the size of the resulting trees and the validated values of evaluation metrics. Throughout this section, we compare our methods to the standard univariate trees of~\cite{leblanc1993survival}, where at each non-terminal node, every possible partition of the data by individual covariates is considered and the split with the largest log-rank statistic is chosen. 

\subsection{Kernels used in Models} \label{sec:COK}

In this paper we implement polynomial kernels
\begin{equation} 
K(\mathbf{u}, \mathbf{v}) = (\mathbf{u} \cdot \mathbf{v} + c)^d
\label{eq:polyker}
\end{equation}
with positive integers $d$ and constant $c \geq 0$, and Gaussian kernels
\begin{equation}
K(\mathbf{u}, \mathbf{v}) = \exp\Big(-\frac{\|\mathbf{u} - \mathbf{v}\|^2}{2\sigma^2}\Big)
\label{eq:GaussKernel}
\end{equation}
 with variances $\sigma^2 > 0$. The polynomial kernel allows splits by polynomial surfaces of degree up to $d$, while the Gaussian kernel allows infinite dimensional splits.

 Among the polynomial kernels, we used the linear kernel with $d = 1, c = 0$, and the quadratic kernel with $d = 2, c = 1$. Our linear kernel recovers the oblique splits of \cite{kretowska}, by using a large enough $\kappa > 0$ in~\eqref{ridgedcfex} to make the influence of the ridge penalty negligible. For the Gaussian kernel, we set 
 \begin{equation}
     \sigma^2 = 
 \frac{1}{n}\sum_{i=1}^n\|\mathbf{x}_i - \overline{\mathbf{x}}\|^2, \label{eq:GaussSigmasqr}
 \end{equation} as an overall measure of variability.  We found that using a  much smaller $\sigma^2$ produces significant overfitting, as illustrated in Figure \ref{fig:ksigmawas} below.

\subsection{Evaluation of Models} \label{sec:evalofmodels}
A tree is trained, $\mathcal{T}_\text{Train}$, on a training subset of the data $(\mathbf{x}_i, t_i, \delta_i)_{i \in \text{Train}}$. At each terminal node, $\mathcal{N}$, of $\mathcal{T}_\text{Train}$, we fit a Kaplan-Meier estimate $S^\ast(t ; \mathcal{N})$. For any observation $(\mathbf{x}_i, t_i, \delta_i)_{i \in \text{Test}}$ denote the Kaplan-Meier estimate of the terminal node to which it belongs as $S^\ast(t ; \mathbf{x}_i)$ and denote the estimated median of  $S^\ast(t ; \mathbf{x}_i)$ as $t_i^\ast$. The predicted survival time with respect to  $\mathbf{x}_i$ from our model is $t_i^\ast$. 

To validate our models, we consider the concordance index and the integrated Brier score. The concordance index quantifies the quality of survival rankings, with a concordance index of 0.50 the expected concordance index when assigning survival times randomly~\citep{brentnall2018use}. It is defined as 
\begin{equation} \label{eq:concord} 
CI_{\text{Test}} = \frac{ \sum\limits_{ \substack{i, j \in \text{Test} \\ t_i^\ast \neq t_j^\ast } } 1_{t_i < t_j} \cdot 1_{t_i^\ast < t_j^\ast} \cdot \delta_i }{\sum\limits_{ \substack{i, j \in \text{Test} \\ t_i^\ast \neq t_j^\ast } } 1_{t_i < t_j} \cdot \delta_i }.
\end{equation}  The Brier score is given as 
\begin{align}
BS_{\text{Test}}(t) = \frac{1}{N_\text{Test}} \sum_{i \in \text{Test}} \Big\{&S^\ast(t ; \mathbf{x}_i)^2 \cdot 1_{t_i \leq t} \cdot \delta_i \cdot [\widehat{G}(t_i)]^{-1} \nonumber \\
    + &(1 - S^\ast(t ; \mathbf{x}_i))^2 \cdot 1_{t_i > t} \cdot [\widehat{G}(t_i)]^{-1}\Big\}
\end{align} where $\widehat{G}(t)$ is the Kaplan-Meier estimate of the censoring distribution. It used to evaluate the accuracy of predicted survival probabilities at a time $t$. Useful models have a Brier score below 0.25, with 0.25 the expected Brier score if $S^\ast(t; \mathbf{x}_i)$ is fixed at 0.50~\citep{ gerds2006consistent}. An overall, or integrated Brier score, is given as 
\begin{align} \label{eq:ibs}
IBS_{\text{Test}} = \frac{1}{\max\limits_{i \in \text{Test}} t_i}\int_0^{\max\limits_{i \in \text{Test}} t_i} BS_{\text{Test}}(t) \ dt,
\end{align} 
which we estimate with the trapezoidal rule over the arithmetic means of uncensored survival times in the testing set. 

\subsection{Models Applied to Simulated Data} \label{sec:simresults}
\subsubsection{Detection of Oblique and Non-Linear Splits} \label{sec:detsplits}
We illustrate our algorithm's ability to accurately split data simulated with non-linear effects on survival times. We sample covariates from 
\[
\mathbf{X} \sim  \mathcal{N}\left[\begin{pmatrix}
1 \\
2
\end{pmatrix},
\begin{pmatrix}
1 & 0 \\
0 & 2
\end{pmatrix}
\right]
\]
with survival and censoring times sampled independently, with respective hazards
\[
h_s(t; \beta_0, \boldsymbol{\beta})= \beta_0 + \langle \boldsymbol{\beta}, \phi(\mathbf{X})\rangle \text{ and } h_c(t; \alpha_0) =\exp(\alpha_0).
\]
$\phi(\mathbf{X})=(X_1,X_2,X_1 X_2, X_1^2, X_2^2)$ is the feature map for quadratic augmentation, $\boldsymbol{\beta}$ is a vector of coefficients and $\beta_0$ and $\alpha_0$ are constants. In each case, we sample $n=180$ observations, with approximately $10\%$ of the data censored in each sample. Censoring comes from the censoring distribution, but also from the survival distribution when an observation is greater than a fixed follow-up time. 

\begin{figure}[H]
\centering
\captionsetup[subfloat]{justification=centering}
\subfloat[Linear Hazard, Linear Kernel]{
\label{fig:LinearFit}
\includegraphics[width=0.3\textwidth]{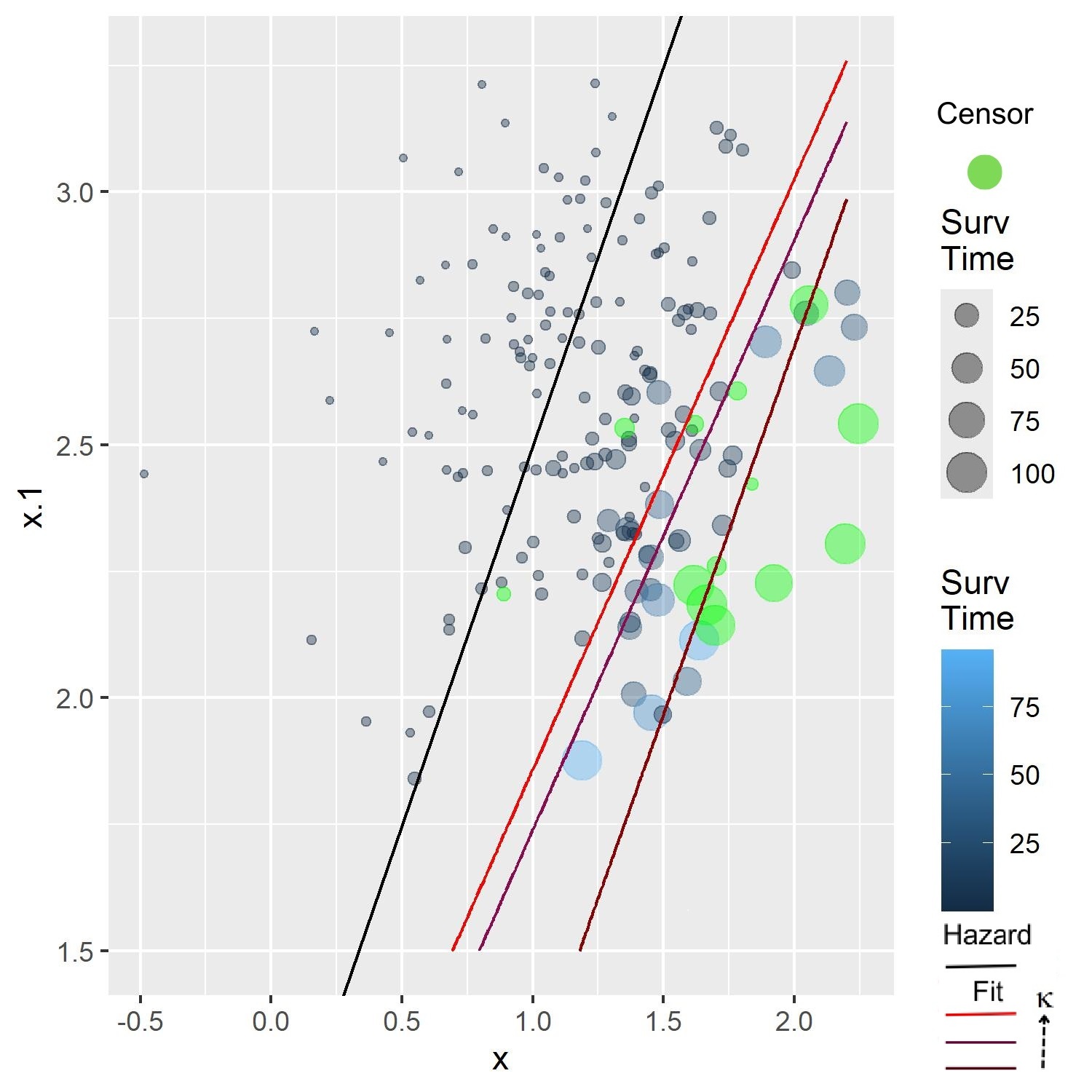}
}
\subfloat[Parabolic Hazard,  Quadratic Kernel]{
\includegraphics[width=0.3\textwidth]{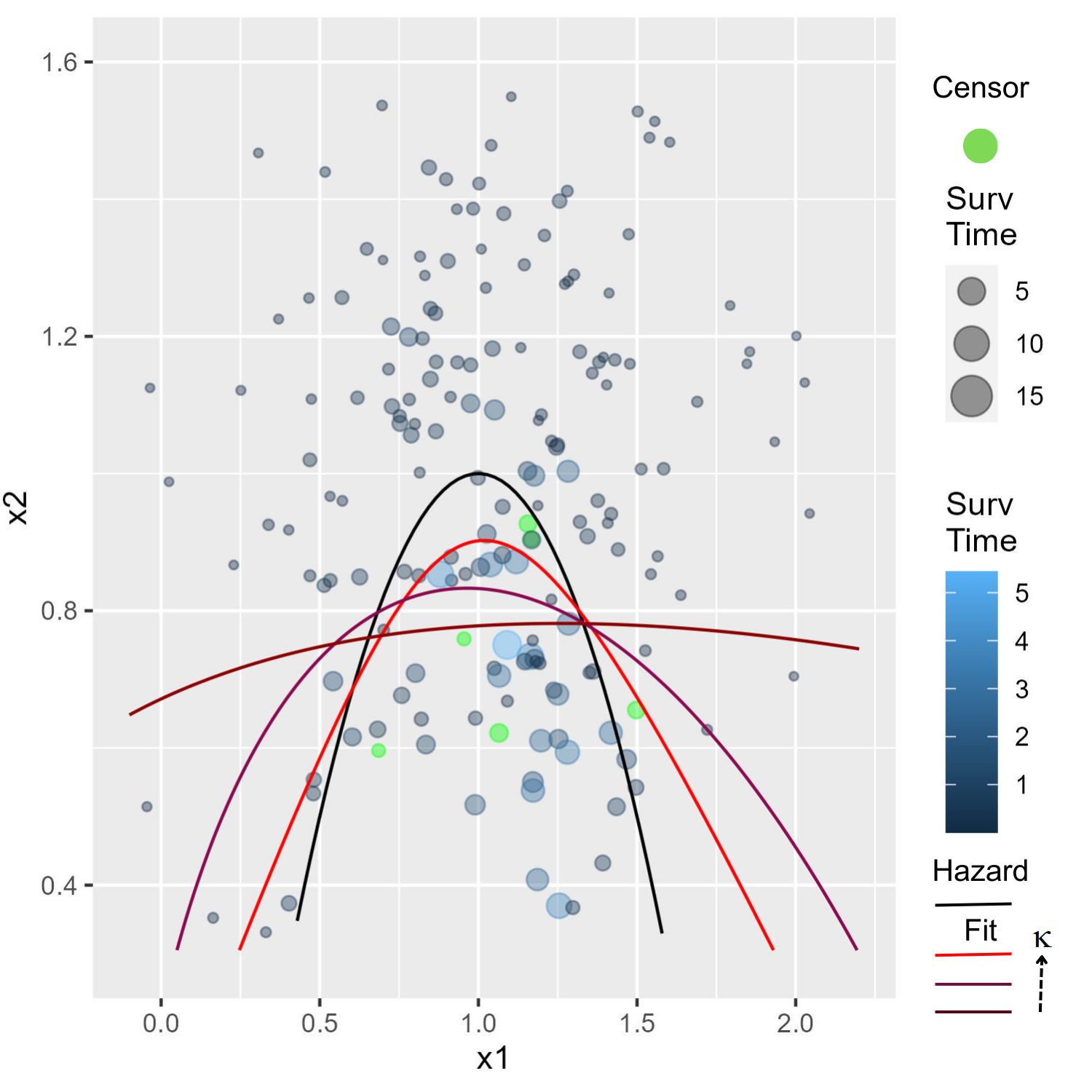}
  \label{fig:ParabolaFit}
}
\subfloat[Elliptical Hazard, Quadratric Kernel]{
\includegraphics[width=0.3\textwidth]{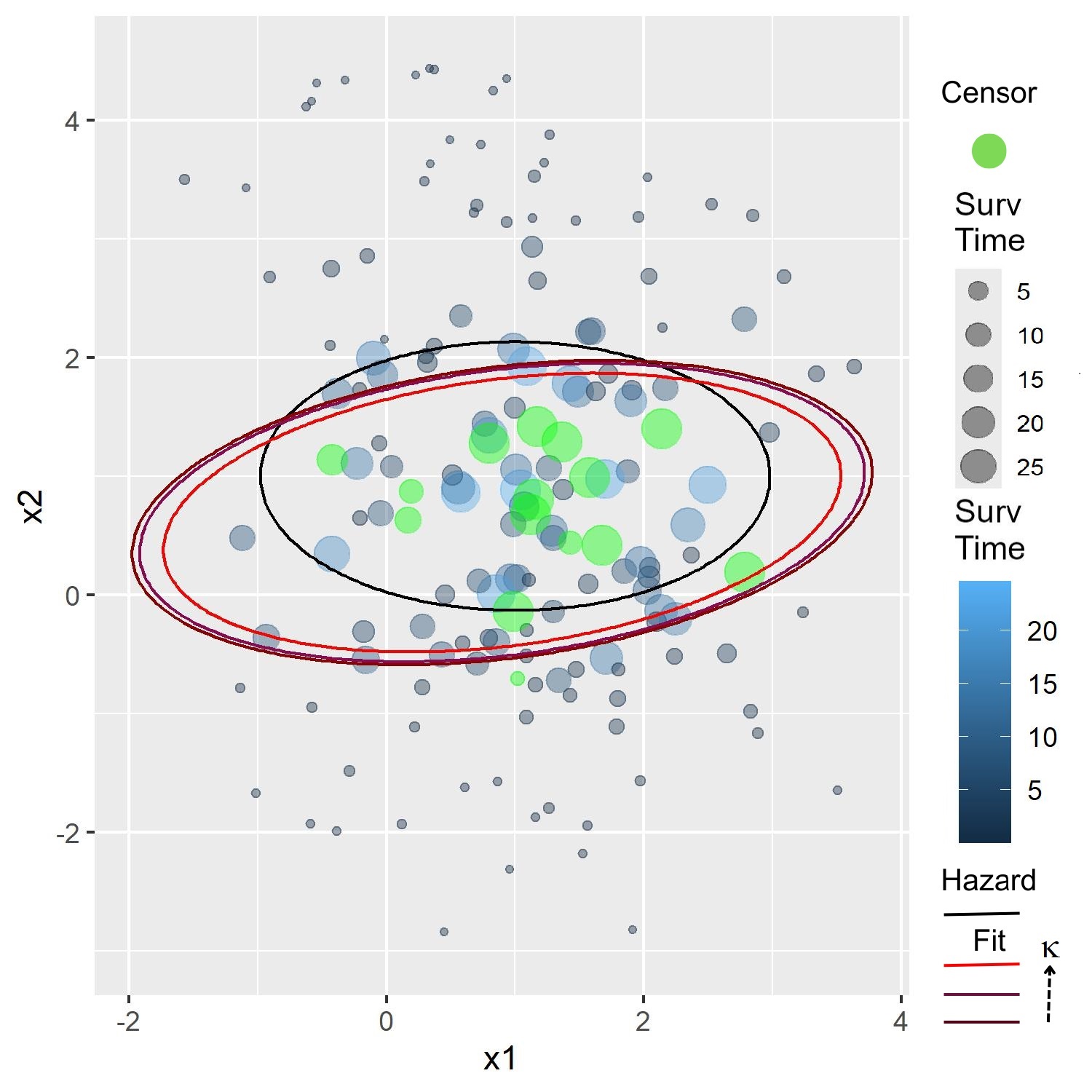}
  \label{fig:EllipticalFit}
}

\subfloat[Hyperbolic Hazard, Quadratic Kernel]{
\label{fig:HyperbolicFit}
\includegraphics[width=0.3\textwidth]{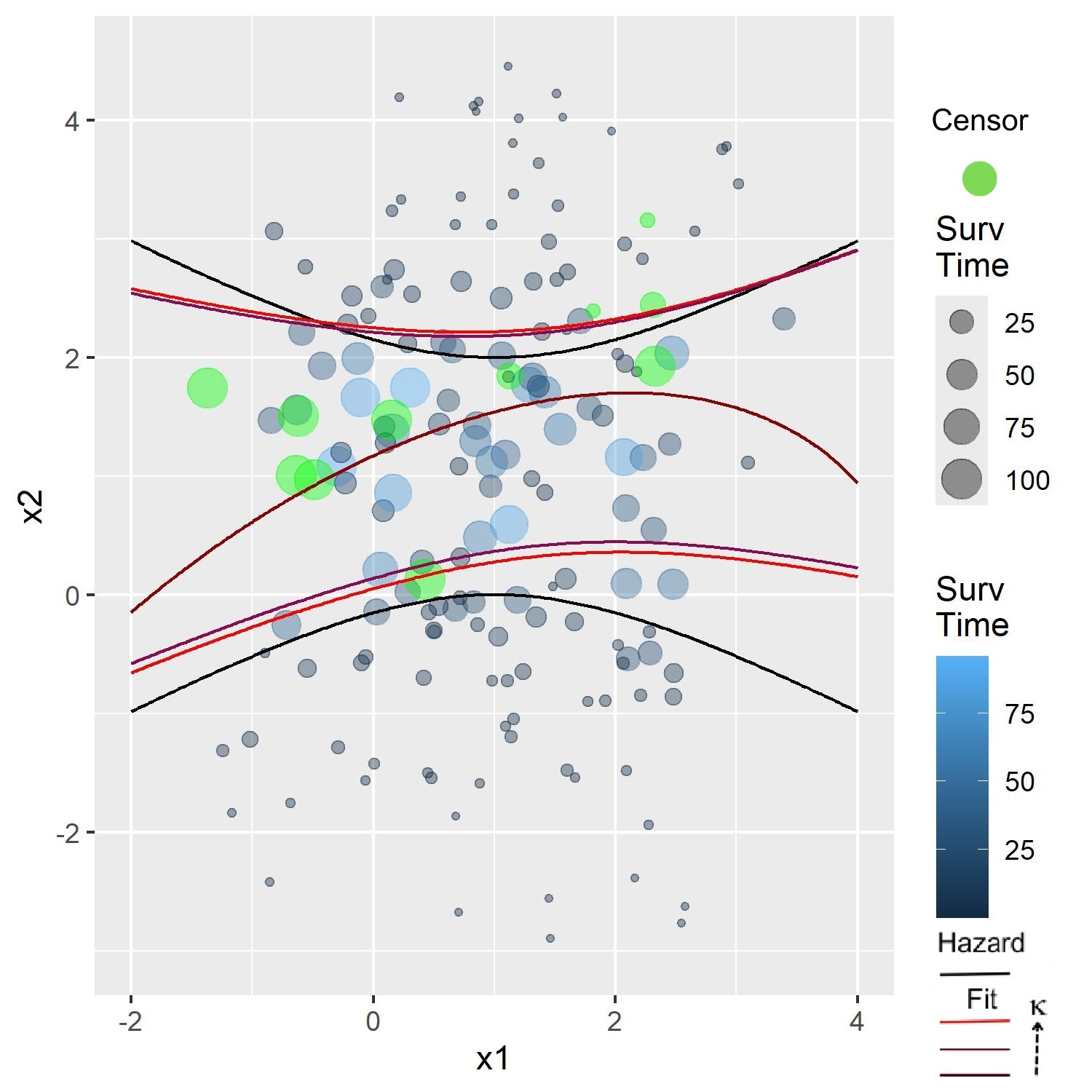}
}
\subfloat[Parabolic Hazard,  Gaussian Kernel]{
\includegraphics[width=0.3\textwidth]{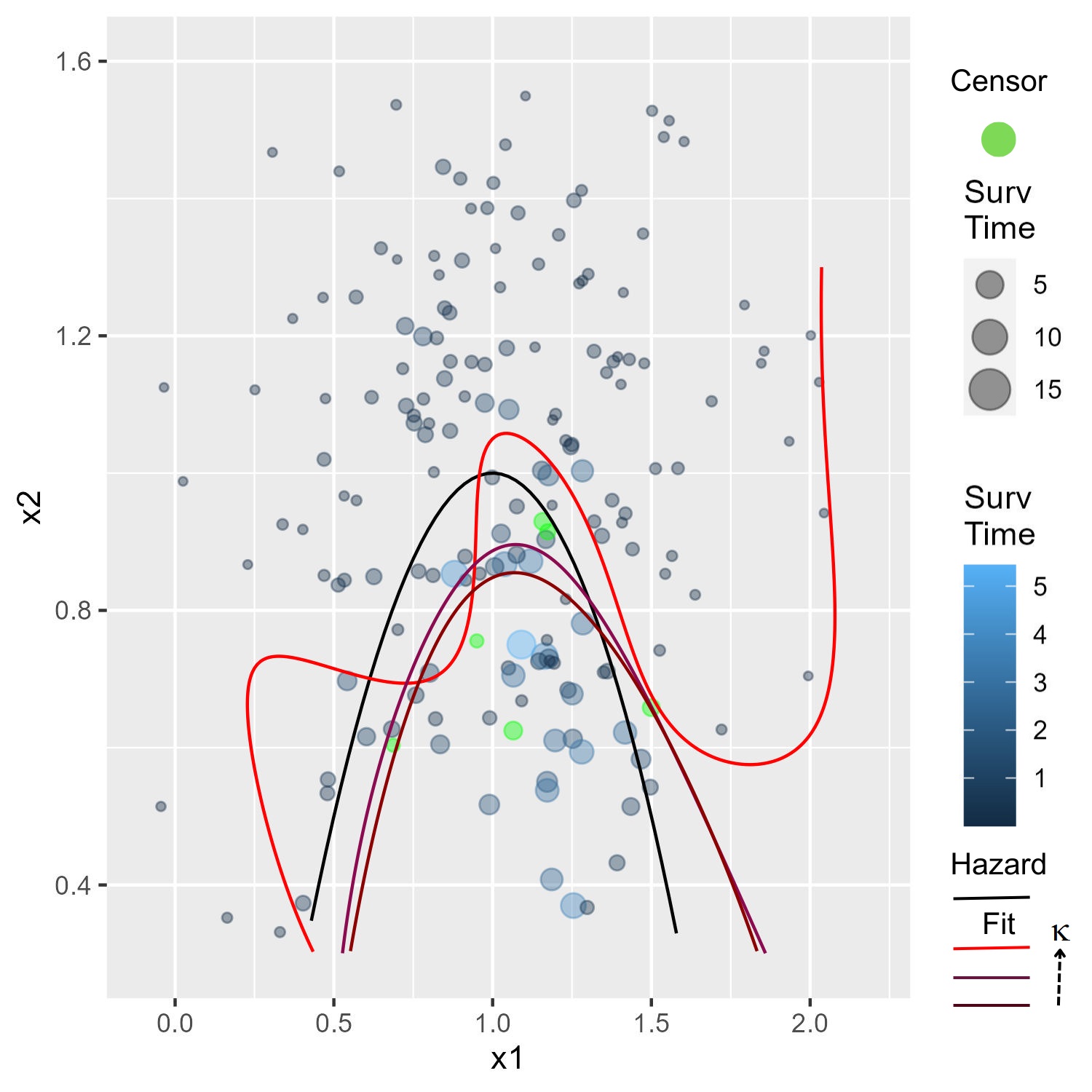}
  \label{fig:ParabolicFitG2}
}
\subfloat[Elliptical Hazard, Gaussian Kernel]{
\includegraphics[width=0.3\textwidth]{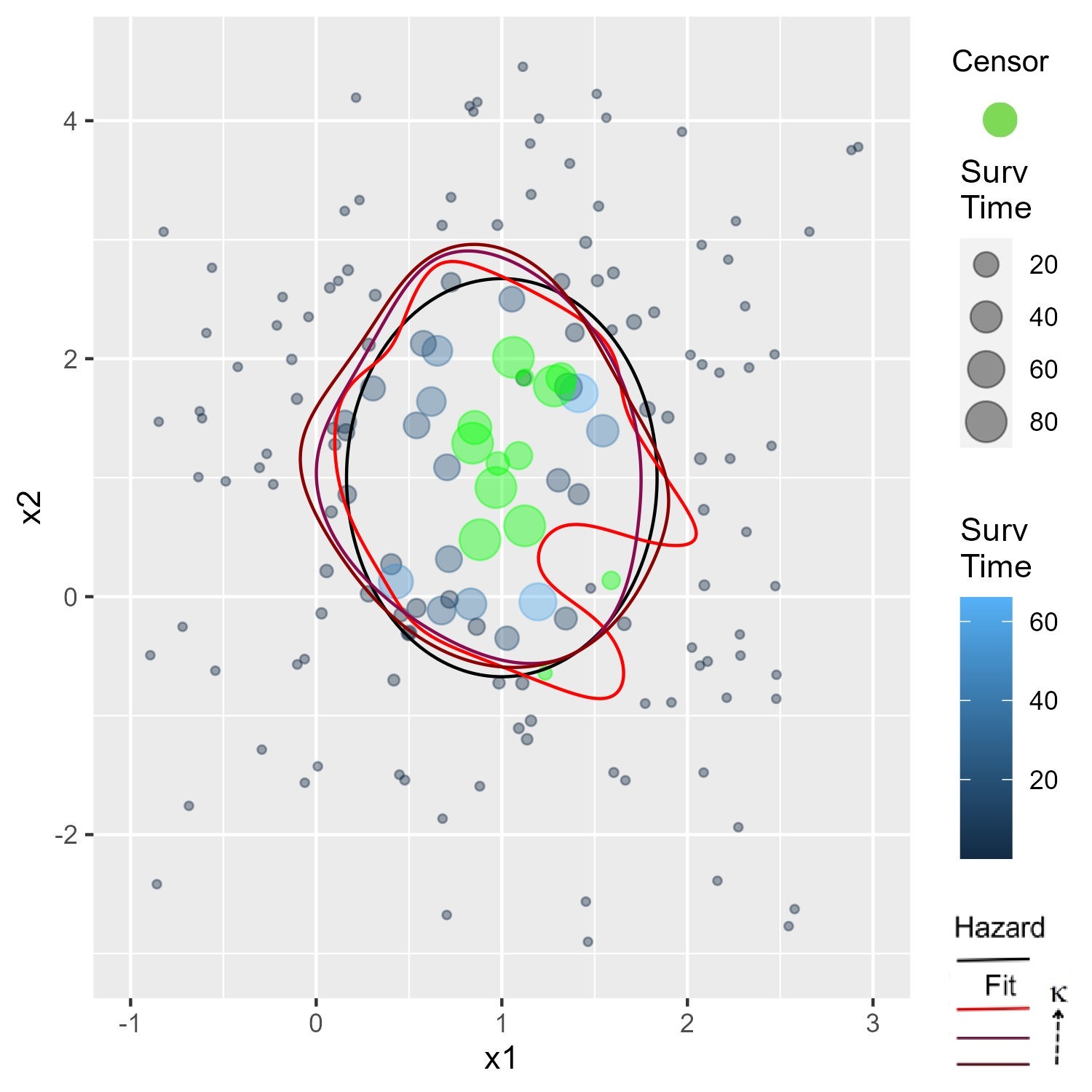}
  \label{fig:EllipseFitG2}
}

\caption{Splits of Data Simulated With Quadratic Hazards }
\label{fig:SplitsofQuadHazards}
\end{figure}

Varying coefficients in $\boldsymbol{\beta}$, in Figure~\ref{fig:SplitsofQuadHazards} we produce oblique and three types of quadratic hazards (parabolic, elliptical and hyperbolic) and plot level hazard sets along with the splitting surfaces located by the algorithm of Section~\ref{sec:ROA}. Using different kernels of section~\ref{sec:COK}, we fit the splitting surfaces using our recursive algorithm~\ref{algo:roo}. 

In Figure~\ref{fig:LinearFit}, we implement a linear kernel of~\eqref{eq:polyker} with $d=1$ and $c=0$. In Figures~\ref{fig:ParabolaFit},~\ref{fig:EllipticalFit} and~\ref{fig:HyperbolicFit}, we implement a quadratic kernel with $d=2$ and $c=1$. In Figures~\ref{fig:ParabolicFitG2} and~\ref{fig:EllipseFitG2}, we use a Gaussian kernel with $\sigma^2$ set as in~\eqref{eq:GaussSigmasqr}. In each figure, we vary the tuning parameter $\kappa$ of ridge regularized splitting function~\eqref{ridgedcfex} to see its effect on the surface fit to the data by the algorithm. With the linear and quadratic kernels, we see splits that are more similar to the hazard for larger the value of $\kappa$. For large values of $\kappa$, the Gaussian kernel produces tight fits to the data which deviate from the hazard. Thus, the importance of tuning $\kappa$ in training, to mitigate overfitting of survival trees, is illustrated. In further simulation, as expected, we observed less accurate fits with larger proportions of censoring. 

\subsubsection{Models Evaluated on Simulated Data} 
\label{sec:modevalsimdata}
In this section, we simulate data in different scenarios from the \texttt{survsim} package~\cite{survsim}. We run 1000 simulations for each scenario. For each simulation, we generate data with $n=500$ observations. On $250$ of those observations, we train and prune the tree model, as well as tune for $\kappa$ in criterion function~\eqref{ridgedcf} using 5-fold cross validation.  The remaining $250$ observations are used to assess the model.

For $p=2$, we generate data as in~\ref{sec:detsplits}, varying $\boldsymbol{\beta}$ so that we have linear, parabolic and elliptical hazards. We also generate data from a Weibull 
distribution with increasing, elliptical hazards. In addition, we run our simulations with $p=4$ and $p=7$ covariates and generate hyper-linear and elliptical hazards. Approximately 10\% to 20\% of the data in each simulated data set is censored.  For each simulation, we fit,  tune and validate trees with the univariate splits of~\cite{leblanc1993survival} and trees with linear, quadratic and Gaussian kernels in algorithm~\ref{algo:roo}. 

Over the 1000 runs, on our testing sets, we compute 1) the overall average Integrated Brier Scores (IBS), taken over the arithmetic means of survival times and 2) the overall average concordance index (CI). The average number of nodes in the trees grown on the training sets (No. Nodes) and the average number of nodes after pruning (No. Nodesp) are also reported. In parenthesis, next to each average, we report the standard deviation of the measurements over the 1000 runs.  

In Table~\ref{tab:sim1}, Table~\ref{tab:sim2} and Table~\ref{tab:sim3}, we fit the univariate (Univar) trees of~\cite{leblanc1993survival}, and trees with our algorithm~\ref{algo:roo} with linear, quadratic (Quad) and Gaussian (Guass) kernels as in Section~\ref{sec:COK}. 

\begin{table}[ht]
\begin{tabular}{  c c c c c c }
 Model & Hazard & IBS & CI & No. Nodes & No. Nodesp \\ \hline \hline
 Univar & Pla &0.084 ($\pm 0.011$) & 0.835 ($\pm0.017$) & 47.58 ($\pm 5.27$) & 18.38 ($\pm 3.66$) \\ 
 Linear & Pla &0.081 ($\pm 0.081$) & 0.941 ($\pm0.018$)  & 7.59 ($\pm 0.81$) & 5.89 ($\pm 0.79$) \\ 
 Quad & Pla &0.069 ($\pm 0.009$)  & 0.924 ($\pm 0.020$)  & 7.91 ($\pm 2.02$) & 5.70 ($\pm 1.85$) \\  
 Gauss& Pla & 0.082 ($\pm 0.010$) & 0.946 ($\pm 0.017$)  & 3.00 ($\pm 0.00$)  & 3.00 ($\pm 0.00$)  \\ \hline
 Univar & Par &0.093 ($\pm 0.010$)  & 0.810 ($\pm 0.024$) & 76.15 ($\pm 5.89$)  & 20.98 ($\pm 3.80$)  \\ 
 Linear & Par &0.108 ($\pm 0.013$)   &0.744 ($\pm 0.042$)  & 7.44 ($\pm 2.10$) & 6.50 ($\pm 1.69$) \\ 
 Quad & Par &0.097 ($\pm 0.011$)  &0.847 ($\pm 0.033$) & 8.92 ($\pm 1.31$) & 4.91 ($\pm 2.30$) \\ 
 Gauss & Par  &0.093 ($\pm 0.010$) & 0.854 ($\pm 0.026$)  & 8.62 ($\pm 1.28$) & 5.41 ($\pm 2.05$) \\ \hline 
 Univar & Ell &0.098 ($\pm 0.011)$ & 0.836 ($\pm 0.035$)  & 56.42 ($\pm 6.17$)  & 17.91 ($\pm 3.71$)\\ 
 Linear & Ell &0.115 ($\pm 0.014$) &  0.620 ($\pm 0.050$) & 14.66 ($\pm  3.81$) & 12.66 ($\pm  3.78$)\\ 
 Quad & Ell  &0.107 ($\pm  0.011$) & 0.789 ($\pm 0.052$) & 13.92 ($\pm 3.10$)   & 4.76 ($\pm 2.81$) \\  
 Gauss  & Ell &0.108 ($ \pm 0.012$) &  0.754 ($\pm 0.063$) & 7.34 ($\pm 1.90$) & 4.79 ($\pm 2.06$) \\ \hline
Univar & Wbl &0.092 ($\pm 0.012$) & 0.840 ($\pm 0.032$)  & 46.56 ($\pm 5.47$) & 19.10 ($\pm 3.20$) \\ 
 Linear & Wbl &0.141 ($\pm 0.014$) &  0.589 ($\pm 0.054$)  & 14.36 ($\pm 4.12$) & 16.89 ($\pm 4.65$) \\ 
 Quad & Wbl &0.082 ($\pm 0.013$)   & 0.821 ($\pm 0.053$) & 11.35 ($\pm 4.21$) & 4.56 ($\pm 3.34 $)   \\  
 Gauss & Wbl &0.081 ($\pm 0.011$) &  0.763 ($\pm 0.056$) & 6.98 ($\pm 1.68 $) & 4.82 ($\pm 2.16 $)\\ \hline
                         \multicolumn{6}{@{}p{4.4in}}{\footnotesize IBS = Integrated Brier Score, \footnotesize CI = Concordance Index}\\
                         \multicolumn{6}{@{}p{4.4in}}{\footnotesize No. Nodes = Number of Nodes in Tree, No. Nodesp = Number of Nodes after Pruning}
\end{tabular}
\caption{Results of Simulations with $p=2$} 
\label{tab:sim1}
\end{table}

\begin{table}[ht]
\begin{tabular}{ c c c c  c c }
 Model & Hazard & IBS & CI  & No. Nodes & No. Nodesp \\ \hline \hline
 Univar & Pla &0.123 ($\pm 0.013) $ & 0.691 ($\pm 0.031$) & 56.22 ($\pm 6.75$) & 12.12 ($\pm 5.21$) \\ 
 Linear  & Pla &0.118 ($\pm  0.012$) & 0.812 ($\pm 0.035$)  & 9.21 ($\pm 2.38$) & 5.36 ($\pm 1.54$) \\ 
 Quad & Pla &0.106 ($\pm 0.011$) & 0.796 ($\pm 0.035$) & 11.11 ($\pm 2.58$) &  6.89 ($\pm 2.17$) \\  
 Gauss& Pla & 0.110 ($\pm 0.012$) & 0.820 ($\pm 0.037$)  & 5.52 ($\pm 1.29$) & 3.33 ($\pm 1.18$) \\ \hline
 Univar & Ell &0.112 ($\pm 0.011$) & 0.663 ($\pm 0.028$)   & 55.24 ($\pm  8.38$)& 36.65($\pm 7.22)$ \\ 
 Linear & Ell &0.132 ($\pm  0.015$) &  0.744 ($\pm 0.045$)  &10.2 ($\pm 2.27) $& 8.40 ($\pm 2.42$) \\ 
 Quad & Ell &0.101 ($\pm 0.013) $& 0.829 ($\pm 0.035$)  & 9.12 ($\pm 1.84$) &7.34 ($\pm 2.45$)  \\  
 Gauss & Ell &0.092 ($\pm 0.013$)&  0.850 ($\pm 0.036$) & 7.31 ($\pm 1.33$) & 3.84 ($\pm 1.66$) \\ \hline
                          \multicolumn{6}{@{}p{4.4in}}{\footnotesize IBS = Integrated Brier Score, \footnotesize CI = Concordance Index}\\
                         \multicolumn{6}{@{}p{4.4in}}{\footnotesize No. Nodes = Number of Nodes in Tree, No. Nodesp = Number of Nodes after Pruning}
\end{tabular}
\caption{Results of Simulations with $p=4$} 
\label{tab:sim2}
\end{table}

\begin{table}[ht]
\begin{tabular}{ c c c  c c c }
 Model & Hazard & IBS & CI  & No. Nodes & No. Nodesp \\ \hline \hline
 Univar & Pla &0.136 ($\pm  0.013$) & 0.651 ($\pm 0.032$)
 & 60.76 ($\pm  6.23$)  & 16.20  ($\pm 5.95$) \\ 
 Linear & Pla &0.121 ($
 \pm 0.009$) & 0.802 ($
 \pm 0.039$)  & 10.52 ($1.948$) & 7.21 ($\pm 2.91$) \\ 
 Quad & Pla &0.079 ($
 \pm  0.009)$ & 0.761 ($\pm 0.040$  & 10.98 ($
\pm  2.68)$ & 8.45 ($\pm 2.43)$ \\  
 Gauss& Pla & 0.102 ($\pm 0.010$) & 0.836 ($\pm 0.040$) & 3.95 ($\pm 1.00$) & 3.12 ($\pm 0.85$) \\ \hline
 Univar & Ell &0.158 ($\pm 0.013) $ &0.630 ($
 \pm 0.029) $  & 53.78 ($\pm  6.72) 
$ &  40.00  ($\pm  8.22$) \\ 
 Linear & Ell &0.138 ($\pm 0.014$) &  0.701 ($\pm 0.056$) & 8.73 ($\pm 2.53$) & 7.30 ($\pm 2.52$) \\ 
 Quad & Ell &0.131 ($\pm 0.015$) &  0.795 ($\pm 0.039$) &   8.86 ($\pm 2.674$) & 6.91 ($\pm 2.75$) \\  
 Gauss & Ell &0.108 ($\pm 0.012$) &  0.792 ($\pm 0.051$) &4.03 ($\pm 1.66$) & 3.76 ($\pm 1.48$) \\ \hline
                         \multicolumn{6}{@{}p{4.4in}}{\footnotesize IBS = Integrated Brier Score, \footnotesize CI = Concordance Index}\\
                         \multicolumn{6}{@{}p{4.4in}}{\footnotesize No. Nodes = Number of Nodes in Tree, No. Nodesp = Number of Nodes after Pruning}
\end{tabular}
\caption{Results of Simulations with $p=7$} 
\label{tab:sim3}
\end{table}

\subsection{Models Applied to Real Data} \label{sec:appresults}
\subsubsection{Remission Data Set} 
\label{sec:remisisondata}
We consider right-censored survival data with $n=42$ observations from~\cite{acute1963effect}, in which the response variable is time until relapse of leukemia patients in remission. A treatment variable indicates whether patients were given a new treatment (TR=0) or a standard treatment (TR=1), with 21 patients given the new treatment.  The other two variables in the data set are log white blood cell count (logWBC) and sex, of which we consider logWBC as it, along with TR, shows significant effect on failure times $(p\text{-value} \approx 0)$ in a standard Cox PH model. 

\begin{figure}[H]
\centering
\begin{subfigure}{.4\textwidth}
  \centering
  \includegraphics[scale=.4]{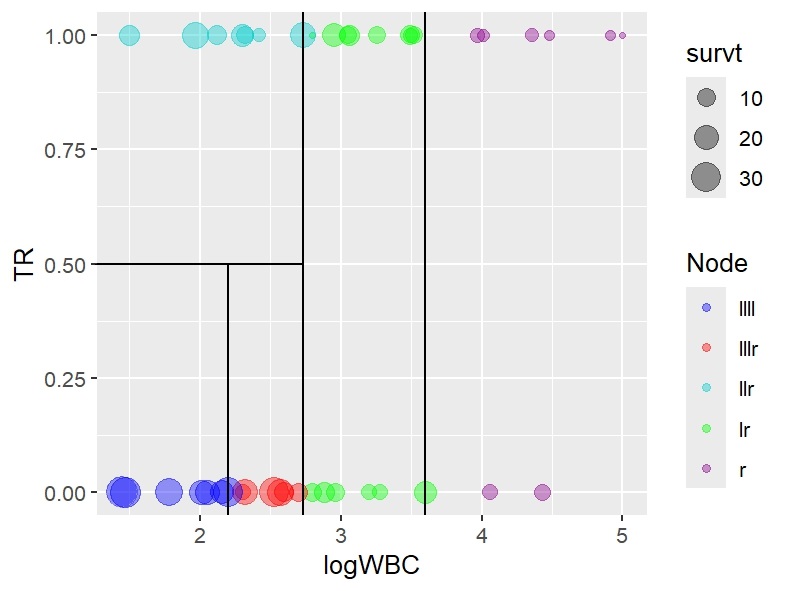}
  \caption{Univariate}
  \label{fig:Unisplit}
\end{subfigure}%
\begin{subfigure}{.4\textwidth}
  \centering
  \includegraphics[scale=.4]{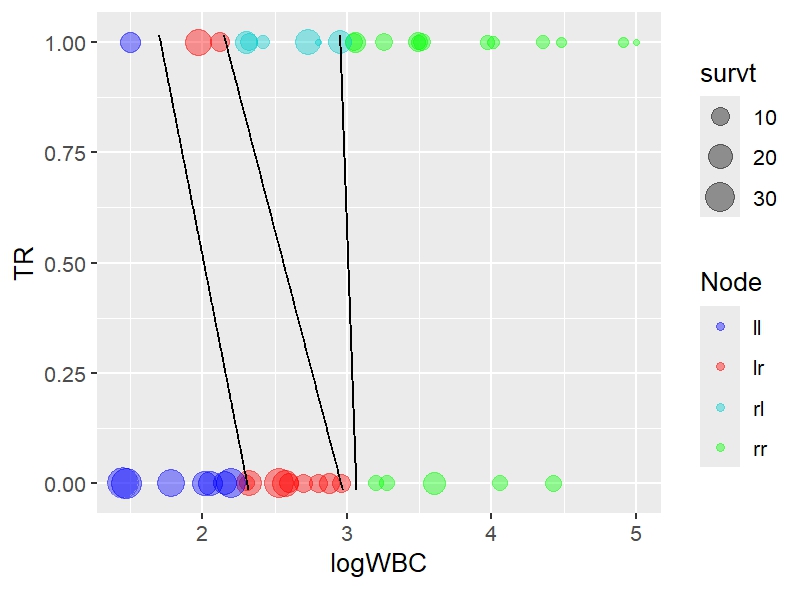}
  \caption{Linear}
  \label{fig:linsplit}
\end{subfigure} 
\begin{subfigure}{.4\textwidth} 
  \centering
  \includegraphics[scale=.4]{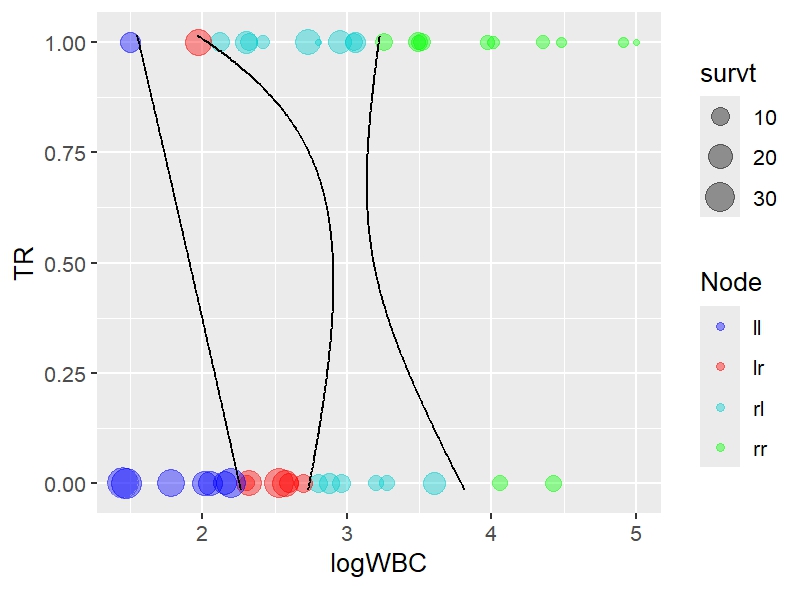}
  \caption{Quadratic}
  \label{fig:quadsplit}
\end{subfigure}%
\begin{subfigure}{.4\textwidth} 
  \centering
  \includegraphics[scale=.4]{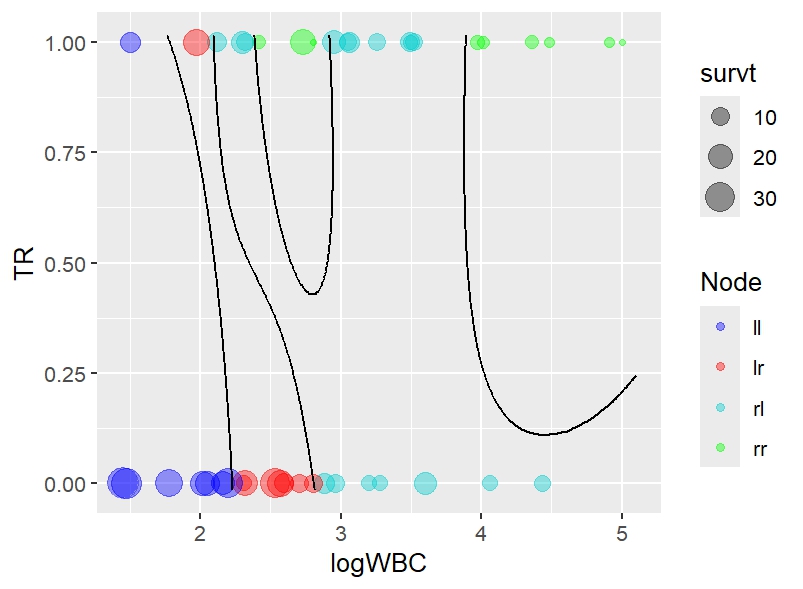}
  \caption{Gaussian}
  \label{fig:gaussplit}
\end{subfigure}

\caption{Tree Models of Remission Data}
\label{fig:RemissionSplits}
\end{figure}

In Figure~\ref{fig:RemissionSplits}, we see the splits in the covariate space, of the terminal nodes of pruned survival trees. These nodes are created through the univariate splits of~\cite{leblanc1993survival} and splits with linear, quadratic and Gaussian kernels as in Section~\ref{sec:COK}.   

Using estimates of survival times as in Section~\ref{sec:evalofmodels}, we have the estimates in Table~\ref{tab:treests} in the terminal nodes. 

\begin{table}[ht]
\begin{subtable}{.23\linewidth}\centering
{\begin{tabular}{c|c}
\ Node & $\hat{T}(\text{Node})$  \\ \hline 
  llll& $26$  \\
  lllr& $23$  \\
  llr & $12$  \\
  lr & $10$ \\
  r & $3.5$ \\
\end{tabular}}
\caption{Univariate}\label{tab:1a}
\end{subtable}%
\begin{subtable}{.23\linewidth}\centering
{\begin{tabular}{c|c}
\ Node & $\hat{T}(\text{Node})$  \\ \hline 
  ll & $24.25$  \\
  lr & $22$  \\
  rl & $11.5$  \\
  rr & $6$ \\
\end{tabular}}
\caption{Linear}\label{tab:1b}
\end{subtable}
\begin{subtable}{.23\linewidth}\centering
{\begin{tabular}{c|c}
\ Node & $\hat{T}(\text{Node})$  \\ \hline 
  ll & $26$  \\
  lr & $23$  \\
  rl & $11.5$  \\
  rr & $8$ 
\end{tabular}}
\caption{Quadratic}\label{tab:1c}
\end{subtable}%
\begin{subtable}{.23\linewidth}\centering
{\begin{tabular}{c|c}
\ Node & $\hat{T}(\text{Node})$  \\ \hline 
  ll & $24.25$  \\
  lr & $23$  \\
  rl & $10$  \\
  rr & $3$ 
\end{tabular}}
\caption{Gaussian}\label{tab:1d}
\end{subtable}%
\caption{KM Median Estimates in Terminal Nodes}\label{tab:treests}
\end{table}
As shown in~\cite{david2012survival}, TR and logWBC meet the assumptions of the Cox PH model. Fitting a Cox PH model with interaction, 
\[
h(t; \boldsymbol{\beta}) = h_0(t) \exp(\beta_1 \, \text{TR} + \beta_2 \, \text{logWBC}  + \beta_3 \, \text{TR} \times \text{logWBC}) 
\]
gives partial maximum likelihood estimates $\hat{\beta}_1 = 2.356, \hat{\beta}_2 = 1.830$  and $\hat{\beta}_3 = -0.342$. The splits in Figure~\ref{fig:RemissionSplits} and the estimates in Table~\ref{tab:treests} agrees with these parameter estimates, with both larger logWBC and TR=1 vs. TR=0 resulting in proportionally larger hazards and, thus, lower survival times. Covariate logWBC has the most significant effect on the hazard and, accordingly, the splits are mostly determined by logWBC in all trees. The estimate  $\hat{\beta}_3 = -0.342$ suggests larger logWBC results in smaller effects of TR, which is also reflected in the splits in Figure~\ref{fig:RemissionSplits} and the survival time estimates in Table~\ref{tab:treests}. The negative slopes of the lines that split the data in Figure~\ref{fig:linsplit} agree with the negative sign of the interaction coefficient, for example.  The splits in Figures~\ref{fig:quadsplit} and ~\ref{fig:gaussplit} most directly take into account interaction and other higher-order effects. Moreover, although the Quadratic tree produces a smaller tree than the univariate splits in Figure~\ref{fig:Unisplit}, it shows better evaluation metrics in Table~\ref{tab:tree_eval} using $k$-fold validation.

\begin{table}[ht] 
\centering
\begin{tabular}{c||l|l|l}
Split & C-Index & IBS  & Nodesp\\ \hline
Univariate  &   0.857 &   0.167 & 8.2    \\
Linear &   0.876 &     0.159 & 7  \\
Quadratic &  0.930 &   0.101  &5.8     \\
Gaussian  & 0.843  &  0.136  & 5   \\
\end{tabular}
\caption{$k$-fold evaluation ($k=5$)} 
\label{tab:tree_eval} 
\end{table}

\subsubsection{Additional Real Data Sets}
\label{sec:addrealdata}
Besides the Remission data set, we consider five real data sets
whose descriptions are given in Table~\ref{tab:realdatasets}. We fit univariate, linear, quadratic and Gaussian trees to training data set aside in $k$-fold cross-validation, and pruned each of these trees according to Section~\ref{sec:prune}. For data sets with $n$ observations, we set $k=10$ for $n<500$ and $k=20$ for $n 
\ge 500$. In Table~\ref{tab:realdatasets_res}, using pruned trees on validation sets, we computed the average integrated Brier scores (IBS), taken over the arithmetic means of survival times, and the average concordance index (CI). We report the the average number of nodes of unpruned trees (No. Nodes), and the average number of nodes of pruned trees (No. Nodesp). In parenthesis, next to each average, we report the standard deviation of these measurements.  
\begin{table}[ht]
\begin{tabularx}{\textwidth}{@{} l|l|l| X @{}}
\toprule
\textbf{Data Set} & $\mathbf{n}$ & \textbf{\% uncensored } & \textbf{Description} \\
\midrule
Addicts
& 239
& 62.76\%
& Australian study from~\cite{caplehorn1991methadone} of time until heroin addicts drop out of methadone treatment. Covariates included in the models were type of clinic (0 or 1), prison record (0 or 1) and methadone dose. \\ \bottomrule
Cost & 518 &  77.99\% & Copenhagen study from~\cite{jorgensen1996acute} of time until stroke of hospital patients with atrial fibrillation and symptoms of stroke as measured by the Neurological Stroke Scale. Covariates included in the models were age, a stroke scale score, presence of diabetes (0 or 1), presence  of atrial fibrillation (0 or 1), and smoker (0 or 1). \\ \bottomrule 
LeukSurv 
& 1043
& 84.28\%
& Study from~\cite{henderson2002modeling} of survival times of acute myeloid leukemia patients in England. Covariates included in the models were age, white blood count at diagnosis, Townsend score measuring neighborhood affluence and spatial statistics of location of residence. \\ \bottomrule 
MGUS
& 241 
& 99.33\%
& Mayo clinic study from~\cite{kyle1993benign} of time in days until diagnosis of a plasma cell malignancy in patients with monoclonal gammopathy. Covariates included in the models were age, sex (M or F), hemoglobin level at diagnosis and monoclonal protein spike size. 
\\ \bottomrule
WHAS500
& 461
& 38.18\%
&  Worcester Heart Attack Study from~\cite{goldberg1988incidence} of survival times of myocardial infarction patients. Covariates included in the models included were inital heart rate, diastolic blood pressure, BMI, gender (0 or 1) and prior history of congestive heart complications (0 or 1).  \\
\bottomrule  
\end{tabularx}
\caption{Description of Real Data Sets}
\label{tab:realdatasets} 
\end{table}

\begin{table}[ht] 
\begin{tabular}{l|lllll}
Data & Model & IBS  & CI         & No. Nodes & No. Nodesp \\
\hline
\multirow{4}{*}{Addicts} & Univ  &  0.178 ($\pm 0.023$)  & 0.658 ($\pm 0.112$)         &   27.4 ($\pm 2.63$)        &     7.6 ($\pm 1.65$)    \\
                         & Lin  & 0.173 ($\pm 0.026$)     & 0.706 ($\pm 0.099$)    &  14.6 ($\pm 1.58$)         &   5.4 ($\pm 0.08$)      \\
                         & Quad  &  0.170  ($\pm 0.025$)   &  0.714 ($\pm 0.102$)            &   16.0 ($\pm 2.71$)         & 6.8 ($\pm 1.03$) \\ 
                         & Gauss &  0.183  ($\pm  0.015$)  &  0.744 ($\pm 0.136$)      &   3.0 ($\pm 0.00$)       & 3.0 ($\pm 0.00$) \\ \hline
\multirow{4}{*}{Cost} & Univ  &  0.192 ($\pm   0.030$)  &  0.624 ($\pm 0.083$)      &  62.40 ($\pm  7.30$)          &     32.0 ($\pm 7.25$)   \\
                         & Lin  &0.178 ($\pm   0.022$)     &  0.740 ($\pm 0.104$)   &  13.6 ($\pm  3.37$)      &  9.2 ($\pm 3.03 $)           \\
                         & Quad  &  0.174 ($\pm 0.021$)   &  0.752 ($\pm 0.186$)         &     13.2 ($\pm 2.96$)      & 8.9 ($\pm 2.93$)   \\ 
                         & Gauss &  0.188 ($\pm 0.027$)  & 0.671 ($\pm 0.215$)     &   6.9 ($\pm 1.22$)      &  5.9 ($\pm 1.02$) \\ \hline

\multirow{4}{*}{LeukSurv} & Univ  &   0.171 ($\pm    0.032$)  & 0.637 ($\pm 0.037$)       &  145.9 ($\pm 10.41$)          &     41.9 ($\pm 10.33$)   \\
                         & Lin  &0.138 ($\pm   0.032 $)     &   0.752($\pm 0.068$)    & 59.2   ($\pm 4.49 $)      &  15.3 ($\pm 4.01 $)           \\
                         & Quad  &  0.154 ($\pm 0.050$)   & 0.760 ($\pm 0.058$)        &     52.0  ($\pm 4.17$)      & 14.4  ($\pm 3.83$)   \\ 
                         & Gauss &  0.142 ($\pm 0.028$)  & 0.758 ($\pm 0.098$)    &    6.2 ($\pm 1.88$)      &   5.6 ($\pm 0.94$) \\ \hline

\multirow{4}{*}{MGUS} & Univ  &  0.167 ($\pm    0.051$)  &  0.646 ($\pm 0.130$)      &      46.8 ($\pm 7.45 $)      &     22.8 ($\pm 6.66$)   \\
                         & Lin  &0.162 ($\pm  0.025$)     &  0.812 ($\pm 0.146$)    &     9.3 ($\pm 2.45  $)      &  6.8 ($\pm 1.93 $)        \\
                         & Quad  &   0.163 ($\pm 0.034$)   &  0.767 ($\pm 0.107$)      &      7.7 ($\pm 2.06$)      &  5.7 ($\pm  1.08$) \\ 
                         & Gauss &  0.164 ($\pm 0.023$)  & 0.761 ($\pm 0.140$)     &  5.5 ($\pm 0.89$)        & 5.0 ($\pm 0.00$) \\ \hline
\multirow{4}{*}{WHAS500} & Univ  &  0.186 ($\pm 0.040$)  & 0.674 ($\pm 0.108$)      &   64.8 ($\pm 5.91$)        &     40.2 ($\pm 4.74$)    \\
                         & Lin  & 0.161 ($\pm 0.044$)     & 0.827 ($\pm 0.128$)    & 22.5 ($\pm 3.51$)      & 9.8 ($\pm 1.93$)          \\
                         & Quad  &  0.165 ($\pm 0.062$)   &   0.836 ($\pm 0.144$)     & 21.6 ($\pm 3.17$)       & 9.5 ($\pm 1.72$)  \\ 
                         & Gauss &  0.166  ($\pm 0.044$)  &  0.871 ($\pm 0.150$)     &   5.0 ($\pm 0.00$)       & 3.7 ($\pm 0.97$) \\ \hline
                         \multicolumn{6}{@{}p{4.4in}}{\footnotesize IBS = Integrated Brier Score, \footnotesize CI = Concordance Index}\\
                         \multicolumn{6}{@{}p{4.4in}}{\footnotesize No. Nodes = Number of Nodes in Tree, No. Nodesp = Number of Nodes after Pruning}
\end{tabular}
\caption{Results on Real Data Sets}
\label{tab:realdatasets_res}
\end{table} 

\begin{figure}[ht]
\centering
 \captionsetup[subfigure]{justification=centering}
\begin{subfigure}{\textwidth}
  \centering
\includegraphics[width=.75\textwidth]{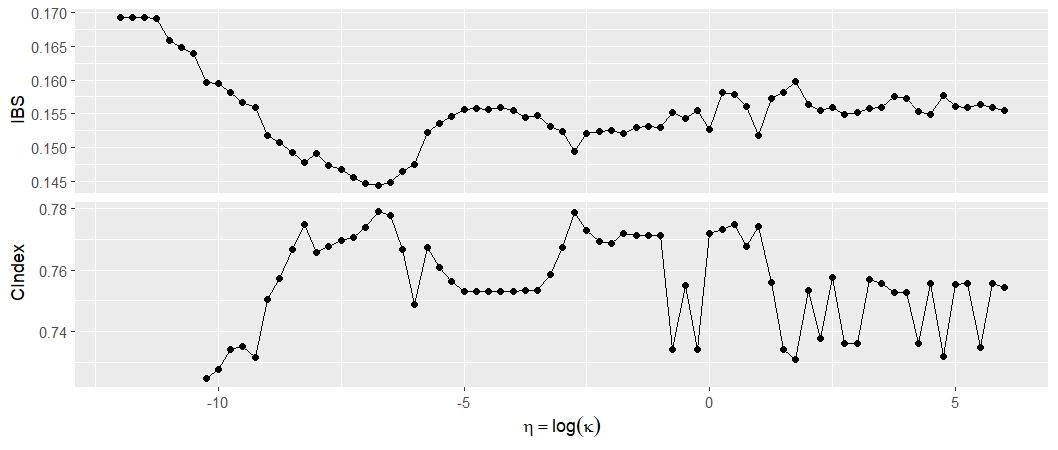}
  \caption{Linear Kernel}
  \label{fig:linchooseeta}
\end{subfigure}
\begin{subfigure}{\textwidth}
  \centering
\includegraphics[width=.75\textwidth]{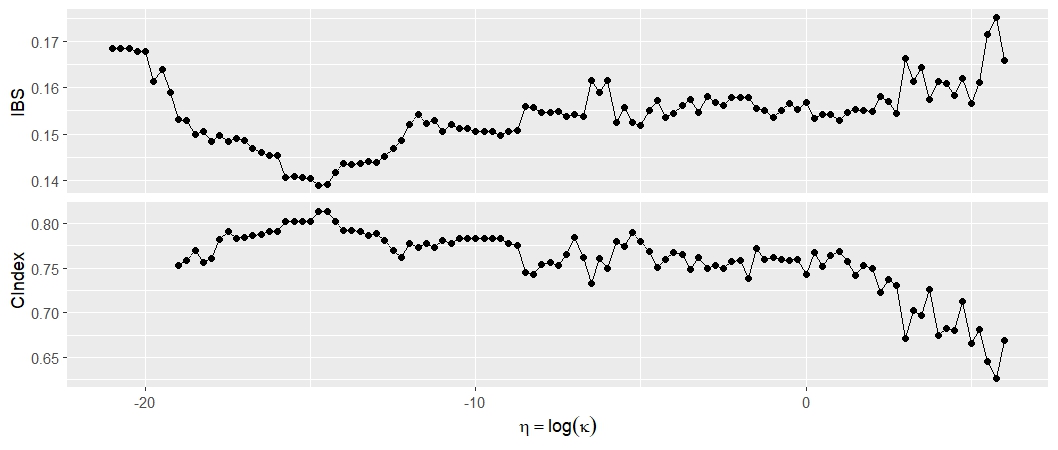}
  \caption{Quadratic Kernel}
  \label{fig:quadchooseeta}
\end{subfigure}
\begin{subfigure}{\textwidth}
  \centering
\includegraphics[width=.75\textwidth]{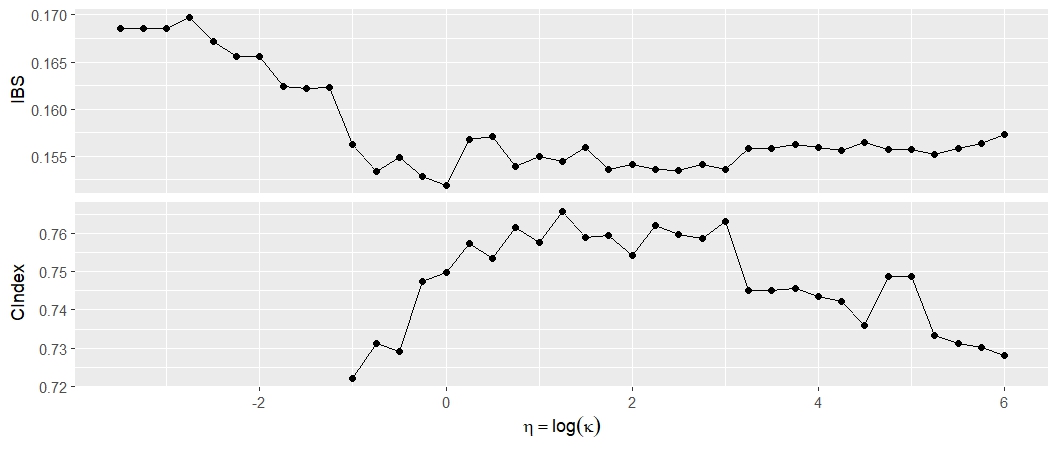}
  \caption{Gaussian Kernel}
  \label{fig:gausschooseeta}
\end{subfigure}
\caption{Cross-validation for $\eta$ on MGUS Data Set}
\label{fig:choose}
\end{figure}

Before fitting each of the models to obtain the results in table~\ref{tab:realdatasets_res}, we used 5-fold validation to choose an appropriate value of $\eta = \log(\kappa)$ in order to determine the regularization parameter $\kappa$. We chose the value of $\eta$ that produced the largest average concordance index over the 5-fold validation runs. We illustrate this process of choosing the tuning parameter on the MGUS data set in Figure \ref{fig:choose}. We observe, as expected, that the average concordance indices are negatively correlated with the average integrated Brier scores. For the linear, quadratic and Gaussian kernels the correlation values observed were $-0.65, -0.86, -0.41$, respectively. We also observe, in each case, the typical shape of a bias-variance tradeoff curve for a regularization parameter like $\kappa$. Indeed, for both small and large values of $\kappa$, the models perform increasingly worse. In particular, for each model, as $\kappa$ becomes very small the survival tree no longer splits to an extent that allows computation of the concordance indices. Similarly, the integrated Brier score converges to a single value that results from its computation on a model with just a single node.  

\begin{figure}[ht]
\centering
  \includegraphics[scale=.4]{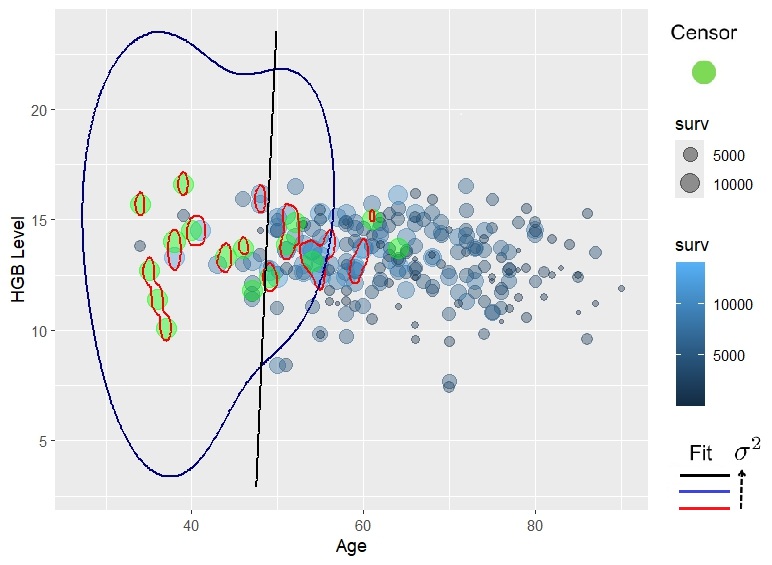}
\caption{Effect of Gaussian Kernel $\sigma^2$ on Splits of MGUS Data Set}
\label{fig:ksigmawas}
\end{figure}

In figure~\ref{fig:ksigmawas}, we see initial splits for the MGUS data set using the Gaussian Kernel. For our validation runs in table~\ref{tab:realdatasets_res}, we use four covariates to split this data. In this figure, however, we use only the two continuous covariates to be able to visualize the split. We see significant overfitting with $\sigma^2=0.25$. The split becomes less complex with larger $
\sigma^2=25$ and nearly linear for $\sigma^2 = 2500$, which is approximately the value of $\sigma^2$ as set in~\eqref{eq:GaussSigmasqr}.

\section{Discussion} 

In Section~\ref{sec:detsplits}, we see that our models are able to locate and approximate linear and non-linear hazard functions. These approximations are done from the sample survival data simulated from distributions with these hazards. We observe the effect of the ridge-regularization parameter in Figure~\ref{fig:SplitsofQuadHazards}. 

In Section~\ref{sec:modevalsimdata}, we simulated data of various dimensions with linear and non-linear hazards. Validating our models, we generally see low integrated Brier scores and high concordance. Compared to the standard univariate trees, the trees induced by our kernelized splitting method are significantly smaller. Furthermore, on the higher dimensional simulations ($p = 4$ and $p = 7$), our method outperformed the predictive power of the univariate trees, in addition to being smaller. The simulations also demonstrate the value of using non-linear splits compared to using only linear splits as in~\cite{kretowska,bobrowskikretowski}. Indeed, trees produced by the Gaussian kernel were consistently smaller than those produced by the linear kernel throughout all simulation cases.

In Section~\ref{sec:addrealdata}, we applied our models to five real data sets, each with right-censored survival responses. Overall, we observed high concordance indices, but somewhat larger integrated Brier scores than in the simulations. The size of the standard univariate trees, were consistently much larger than the trees induced by our method and generally had lower concordance indices and higher Brier scores. Overall, there was significant variability in the results across the folds in the validation process, which is typical for tree models applied to real data sets \citep{joshi2014improving,seni2010ensemble}. Nevertheless, we observed the same pattern as seen in the simulations of Section~\ref{sec:modevalsimdata}, wherein trees produced by the Gaussian kernel were consistently smaller, while preserving similar predictive power.

In summary, we addressed an ambiguity in the assignment of dipole penalty functions in~\cite{kretowska}, by rigorously defining dipole orientation in Section~\ref{sec:OCDPF}. We then regularized the splitting objective function used by~\cite{kretowska}. This allows the application of a wide class of kernels to the splitting function, as detailed in Section~\ref{sec:dualkernel}, and thus enables a variety of non-linear splits for survival trees. This led to the algorithm of Section \ref{sec:ROA}, that recursively outputs splitting vectors and reorients dipoles relative to these vectors until convergence. Finally, we induced survival trees through this splitting algorithm, and we validated this approached on real and simulated data sets in Section~\ref{sec:results}. 

\section{Future Work} 

The structure and size of trees from training depends on parameters such as the percentile cutoffs, $\zeta_1$ and $\zeta_2$, for determining pure and mixed dipoles of Section~\ref{subsec:SDC}, the $\alpha_{jk}$ pure and mixed penalty factors in the dipolar criteria function~\eqref{eq:DPC} of Section~\ref{sec:OCDPF} and the largest allowed size of terminal nodes. Suitable approaches to tuning and setting the values of these parameters can be found by validation. Further, the high variability in validation on real data sets can be addressed by incorporating our trees into various ensemble methods as in~\cite{kretowska2014comparison, krketowska2007ensembles}.

Similar to the SVM methods for classification \cite{Cortes1995}, our method can create efficient models for data with a small number of observations relative to a large number of covariates. This is because the size of kernel matrix only scales with the number of observations. Kernelization and the ridge-regularization are well-known approaches to help alleviate the curse of dimensionality. 

It is not surprising that the trees induced by non-linear kernels, particularly the Gaussian kernel, pruned to smaller trees as these models are able to split data with more flexibility and capture more information sooner in the iterative tree growing process. This suggests a promising direction, as our approach is readily adaptable to other types of non-linear kernels. This includes models that implement polynomial kernels of any degree, logistic kernels, sigmoid kernels and many others.  Our approach is a significant generalization of~\cite{kretowska}. It provides a wide-class of flexible models that can effectively accommodate many types of data with non-linear effects of the covariates on the survival time response. 

In conclusion, our approach allows for flexible and effective tuning, incorporation into ensemble methods and applications to high-dimensional data sets, along with the use of an wide-array of kernels.  Future work can use the basis that we have established in the paper for powerful approaches to modeling right-censored survival data. 

\label{sec:discussion}

\bibliography{Sources}

\end{document}